\documentclass[aps,preprint]{revtex4}
\usepackage{graphicx,amsmath,amssymb}
\usepackage{bm}
\usepackage{xcolor}

\newcommand{\bea}{\begin{eqnarray}}
\newcommand{\eea}{\end{eqnarray}}
\newcommand{\be}{\begin{equation}}
\newcommand{\ee}{\end{equation}}

\numberwithin{equation}{section}

\newcommand{\bfx}{{\bf x}}
\newcommand{\eps}{\epsilon}
\newcommand{\kB}{k_{\rm B}}
\newcommand{\pd}{\partial}

\newcommand{\set}[1]{\{ #1\}}
\newcommand{\braket}[1]{\left\langle #1\right\rangle}
\newcommand{\rttensor}[1]{\overline{\overline{#1}}}

\begin{document}
    
\begin{titlepage}
        
        \title{Hydrodynamic Equations for a system with translational and rotational dynamics}
                
        \author{Akira Yoshimori} 
        \affiliation{Faculty of Science, Niigata University, Niigata, 950-2181, Japan.}
       \author{Shankar P. Das}
        \affiliation{School of Physical Sciences, Jawaharlal Nehru University, New Delhi 110067, India.} 
        
\begin{abstract}
    
     We obtain the equations of fluctuating hydrodynamics for many-particle systems whose microscopic units have both translational and rotational motion. The orientational dynamics of each element are studied in terms of the rotational Brownian motion of a corresponding fixed-length director ${\bf u}$. We consider the microscopic dynamics for two choices of basic variables: Smoluchowsky dynamics for position, Fokker-Planck dynamics for position, and momentum. The time evolution of a set of collective densities $\{\hat{\psi}\}$ is obtained as an exact representation of the corresponding microscopic dynamics. For the Smoluchowski dynamics, noise in the Langevin equation for the director ${\bf u}$ is multiplicative. We obtain that the equation of motion for the collective number-density has two different forms, respectively, for the  I\"{t}o and Stratonvich interpretation of the multiplicative noise in the ${\bf u}$-equation. Without the ${\bf u}$ variable, both reduce to the Standard Dean-Kawasaki form. Next, we average the microscopic equations for the collective densities  $\{\hat{\psi}\}$ (which are, at this stage, a collection of Dirac delta functions) over phase space variables and obtain a corresponding set of stochastic partial differential equations for the coarse-grained densities  $\{\psi\}$ with smooth spatial and temporal dependence. For averaging, we use a general local-equilibrium distribution involving an extended set of dynamical variables for rotational motion. The coarse-grained equations of motion for the collective densities $\{\psi\}$ constitute the fluctuating non-linear hydrodynamics for the fluid with both rotational and translational dynamics. From the stationary solution of the corresponding Fokker-Planck equation, we obtain a free energy functional ${\cal F}[\psi]$ and demonstrate the relation between the ${\cal F}[\psi]$s for different levels of the FNH descriptions with its corresponding set of $\{\psi\}$.

\end{abstract}

        
        \maketitle
    \end{titlepage}
    \setcounter{equation}{0}
    \section{Introduction}
    \label{s1}
    For a system with many microscopic variables, the continuum field theoretic description provides a useful tool for understanding its thermodynamic and time-dependent behaviour. The field theoretic models had been formulated in terms of a set of local densities that can occur due to various reasons like underlying conservation laws \cite{forster}, breaking of continuous symmetry\cite{nambu,goldstone} or presence of some characteristic parameter (like the large inertia of a Brownian\cite{einstein} particle) for the system. These local densities are coarse-grained functions obtained from microscopically defined entities and have smooth space-time dependences.
    Analysis using the equivalent (to the Langevin description\cite{chandra} of FNH) probabilistic approach of the Fokker-Planck equation shows that the stationary distribution for the system is obtained as $\exp[-{\cal{F}}]$, where the free energy  ${\cal F}[\psi]$ is a functional of the coarse-grained variables $\{\psi\}$. Continuum models of the supercooled metastable liquid ( without any quenched disorder) constructed along these lines have been used to study its configurational entropy \cite{remi,wolynes,pre22}. With the assumption of wide separation of characteristic time scales,  local densities are treated as slow variables, and their respective equations of motion are obtained in the form of stochastic PDEs. The stochastic equations\cite{risken} for the local densities constitute the fluctuating nonlinear hydrodynamics (FNH) \cite{skma,ma-mazenko} description. The nonlinear stochastic PDEs for the time evolution of the local densities are treated as plausible generalizations of conventional hydrodynamics. They are useful in studying the effects of cooperativity on the dynamic properties of the system, like relaxation behaviour of correlation of fluctuations and dynamical phase transitions \cite{dm,dm-2009,spd-book,abl,neeta-pre}.
    
    The underlying conservation laws for the studied system often dictate the choice of local densities in the field-theoretic models. 
    For an isotropic liquid, the obvious slow modes are the densities of conserved quantities like mass, momentum, and energy. We focus on mass and momentum densities $\hat{\rho}$ and $\hat{\bf g}$. At the microscopic level, those local densities are expressed in terms of the Dirac delta function involving the position and momentum coordinate $\{{\bf r}_N,{\bf p}_N\}$ (see Eqn. (\ref{def-rho})) for example). The corresponding coarse-grained densities $\{{\rho},{\bf g}\}$ are obtained by averaging the microscopic coordinates using the probability distribution for an appropriate ensemble. When the theoretical description involves both the position and momentum of the constituent particles, it is referred to here as Fokker-Planck dynamics. A simpler formulation in terms of the position variables $\{{\bf r}_N\}$ only is termed the Brownian dynamics. In the latter case, the single equation for the conserved variable density $\rho({\bf x},t)$ with noise  $\theta$ is obtained in the form. 
    \be \label{dk-eqn}
    \frac{\pd{\rho}}{\pd t}
    = D_0 \nabla_{\bf r}{\cdot} {\Big \{}
    {\rho}\nabla_{\bf r} \frac{\delta{F}[{\rho}]}{\delta{\rho}}\Big \}
    {+} {\theta}~.
    \ee
    $D_0$ is a bare diffusion constant related to the strength of the multiplicative noise $\theta$.
    Several methods have been used to obtain this generalized diffusion equation (\ref{dk-eqn}) for $\rho(\bfx,t)$. In one approach, starting from the FNH equations for the extended set $\{\rho,{\bf g}\}$, the momentum density ${\bf g}$ is eliminated\cite{munakata94} by making the so-called adiabatic approximation which assumes that momentum relaxes fast compared to density in the dense liquid. The resulting equation of motion for $\rho({\bf x},t)$ is obtained as the nonlinear Langevin equation in the form (\ref{dk-eqn}). Subsequently, Kawasaki and Miyajima (1998) obtained the same equation from a field-theoretic analysis of the FNH equations for  $\{\rho,{\bf g}\}$ by integrating out the momentum density  ${\bf g}$ and adopting the same adiabatic approximation. A different scheme of reaching the equation for the microscopic density $\hat{\rho}({\bf x},t)$ focused on the microscopic dynamics of the particles. An exact representation of the Langevin dynamics for the position coordinates $\{{\bf r}_\alpha\}$ ($\alpha=1,...N$) of the Brownian particles is a diffusive equation for  $\hat{\rho}({\bf r},t)$, similar to Eq. (\ref{dk-eqn}). The corresponding functional $F[\hat{\rho}]$ in this case is obtained in terms of the {\em bare interaction} potential between the Brownian particles. This equation for $\hat{\rho}({\bf r},t)$ is essentially a balance equation signifying particle conservation. For an extended set of collective variables $\{\hat{\rho},\hat{\bf g}\}$, the corresponding balance equations are put \cite{nakamura,ay0} in the form similar to the equations of FNH, but expressed in terms of functions involving microscopic particle coordinates. The collective densities for which the equations of motion are obtained are a sum of Dirac Delta functions \cite{furusawa,Kalidas}
    The balance equations for the collective densities (signifying the conservation laws for the respective systems) in both Fokker-Planck and Brownian dynamics have been averaged over \cite{pre2013} the local equilibrium ensemble \cite{zuberav} to obtain the coarse-grained equations of FNH. The latter, with smooth space-time dependences, also introduces the free energy functional F[$\rho$], which is shown to be identical to the one used in classical DFT and is expressed in terms of direct correlation functions. For the Brownian case, it was shown \cite{pre2013}, the coarse-grained density obtained \cite{pre2013} is the Eqn. (\ref{dk-eqn}).
    
    The equations of FNH obtained correspond to either the Brownian dynamics or the Fokker-Planck dynamics, which are stochastic PDEs with noise. Here, Eq. (\ref{dk-eqn}) is a simple example (for the case of the Brownian dynamic). Similar but fully deterministic equations without the noise term have been used in literature as the basis of dynamic density functional theory (DDFT) \cite{penna,marconi}. This DDFT equation for $\rho$ has been obtained \cite{archer,lowen} starting from the time evolution of the $N$ particle distribution function ${\cal P}({\bf r}_1...{\bf r}_N)$. Time evolution of ${\cal P}$ is given by the Fokker-Planck equation \cite{vankampen}  corresponding to the Langevin equations for the positions $\{{\bf r}_1,..,{\bf r}_N\}$ of the Brownian fluid. The equation of motion for the one-particle distribution function $\rho(\bfx,t)$ follows from that of the  N-particle distribution function by integrating the coordinates of the other $N-1$ particles. Applying the adiabatic approximation here obtains an equation similar to  Eqn. (\ref{dk-eqn}), without noise.

    A key motivation for studying the field-theoretical models is understanding with theoretical models the role of nonlinear and nonlocal effects in complex systems with many degrees of freedom. Thermodynamics properties come from non-Gaussian couplings (of the fields) in the system's effective free energy functional \cite{KW_rmp}. 
    Working with the stochastic equations for the time evolution of the local fields allows us to compute their correlation functions by averaging over the noise instead of having to average over initial configurations (as is done for Newtonian dynamics \cite{jsp1,jsp2} fluids). The correlation function calculations often adopt diagrammatic methods \cite{msr,msr-pif,msr-pif1} to compute the effects of nonlinearities in the FNH. Equations of fluctuating nonlinear hydrodynamics have been used in many problems to account for nonlinear and nonlocal effects in a complex system. Examples are turbulence in randomly stirred fluid \cite{fns}, dynamic transition in a supercooled liquid \cite{dm,spd90}, dynamics of active matter, \cite{toner-tu,pre2018} Colloids\cite{arXiv2402}, biophysical problems like Cell chemotaxis \cite{goldstein,keller-siegel}, and many more.
    
    In the present work, we consider the formulation of the field-theoretic equations of FNH for a system in which the translational and the orientational degrees of freedom of the individual elements in the fluid are present. The rotational motion for the fluid particles is now included at the microscopic level while formulating the FNH equations. The angular momentum density $\ell({\bf x},t)$ is added to the set $\{\rho,{\bf g}\}$ fields \cite{dokalam,forster-pen,dahler,angmom-nature}. 
    The corresponding driving free energy functional ${\cal F}[\rho,g,\ell]$ appearing in the FNH equations has extra contributions which link to the angular momentum density $\ell$. Like the translational case, there is also an over-damped approximation in this case, which implies setting the time derivative of $\ell$ to zero. The corresponding motion of the orientational vector ${\bf u}$ is precision around the angular velocity $\bm{\omega}$.
    An interesting aspect of this dynamics is that the Langevin equation\cite{langevin} for the rotational Brownian motion now has multiplicative noise\cite{aron,cugli-mnoise}, which requires a proper interpretation of the corresponding stochastic integrals, {\em, i.e.}, adopting either the I\^{t}o\cite{ito} or the Stratonovich interpretation \cite{staton} of the noise. In the present paper, we obtain the Dean-Kawasaki {kawasaki,dean} equation (\ref{dk-eqn}) corresponding to the two interpretations of the multiplicative noise. 
    The paper is organized as follows: In the next Section, we describe the microscopic dynamics of the constituent particles, including both translational and orientational variables; we also introduce here the collective densities in terms of which the equations of fluctuating hydrodynamics are formulated. Section 3 discusses the equations of motions for the collective densities as an exact representation of the microscopic dynamics. Section 4 discusses the local equilibrium distribution in which the coarse-graining of the microscopic equations obtained in Section 3 is done. In Section 5, we obtain the coarse-grained equations for FNH. The results are obtained for both Smoluchowsky and Fokker-Planck dynamics. In Section 6, we analyze the free energy functionals that enter the FNH equations and demonstrate how they are related. We end the paper with a brief discussion of the implications of our results.

\section{Microscopic Dynamics}

We consider a Classical system for which the microscopic dynamics are described in terms of the equations of motion for the position and momentum of its constituent particles: $\{{\bf r}^\alpha,{\bf p}^\alpha\}$ for $\alpha=1,...N$. The microscopic dynamics also includes the rotational motion of the constituent particles in addition to their translational motions described with $\{{\bf r}^\alpha(t),{\bf p}^\alpha(t)\}$. 
The linear velocity of the particle $\alpha$ is ${\bf v}^\alpha$, {\em i.e.}, ${\bf p}^\alpha=m{\bf v}^\alpha$, m being the mass of the particle.
We adopt the notation in which Greek
indices correspond to particles while Roman indices
represent Cartesian components. The repeated Roman indices are understood to be summed over (Einstein convention). 
To represent the orientational motion of the constituent particles,
the rotational dynamics of particle $\alpha$ is considered  in terms of a corresponding director ${\bf u}^\alpha$.  The length $u_0$ of the director is fixed, and its orientation is described in terms of the angles $\{\theta^\alpha,\phi^\alpha\}$ with a set of body fixed axes for the particle  $\alpha$. The origin of the directional vector ${\bf u}$ is placed at the molecule's centre of mass $\alpha$. The microscopic variables describing the motion (translational and rotational) of a single unit (labelled as $\alpha$) are, therefore, a combination of the two vectors $\{{\bf r}^\alpha,{\bf u}^\alpha\}$ with ${\bf u}^\alpha$ being a fixed length vector precessing with angular velocity $\bm{\omega}^\alpha$ around an axis passing through the centre of mass. Thus ${\bf u}^\alpha{\equiv}\{u_0,\theta^\alpha(t),\phi^\alpha(t)\}$. 
In this article, to avoid cluttering, the full set of variables (respectively signifying translational and rotational motions) $\{{\bf r}^\alpha,{\bf u}^\alpha\}$ for element $\alpha$ will be denoted by $\{{\bf x}^\alpha\}$.

In the present work, the interaction energy $U$ for the $N$ particle system is assumed to be dependent on configuration $\{{\bf x}^\alpha\}$ of the particle $\alpha$ ($=1,...N$) which includes both translational and orientational coordinates.
We simplify the analysis by assuming that the total potential energy $U$ is obtained as a sum of pairwise and single particle contributions, denoted by $U_0$ and $U_1$. 
\be
\label{pe1}
U=U_0+U_1~.
\ee
$U_0$ is the pairwise sum of interaction potential $\Phi_0({\bf x}^\alpha-{\bf x}^\nu)$ between two molecules at ${\bf x}^\alpha$ and ${\bf x}^\nu$:
\be
\label{ipote}
U_0{=}\frac{1}{2}\sum_{\alpha,\nu=1}^{{N}\ \prime} 
\Phi_0({\bf x}^\alpha-{\bf x}^\nu)~,
\ee
where the $\prime$ on the summation indicates $\alpha{\ne}\nu$. 
The one body term is obtained as $U_1{=}\sum_{\alpha}^{N} 
\Phi_1({\bf x}^\alpha)$.

To model the director ${\bf u}$ introduced above to signify orientational motions of the particle $\alpha$, the latter is treated as a collection of rigidly connected $s$ units with respective coordinates ${\bf r}_\nu^\alpha$, and mass $m_\nu^\alpha$, for $\nu=1,...s$. Hence in this model, the set $\{{\bf r}^\alpha,{\bf p}^\alpha\}$  for particle $\alpha$ correspond to the centre of mass for the set of elements at $\{{\bf r}^\alpha_\nu,{\bf p}^\alpha_\nu\}$ with $\nu=1,...s$.   
\be \label{cm-defn} {\bf r}^\alpha=\frac{1}{m^\alpha} \sum_{\nu=1}^s
m_\nu^\alpha {\bf r}_\nu^\alpha~, \ee
where $m^\alpha=\sum_{\nu=1}^s m_\nu^\alpha$ is the mass of the $\alpha$-th molecule ($\alpha=1,...N$).
For the $\alpha$-th molecule, coordinates of the constituent elements $\nu$
(for $\nu=1,..s$) are $\{{\bf r}^{\prime\alpha}_\nu\}$ 
with  origin at the centre of mass ${\bf r}^\alpha$ for the $\alpha$-th molecule. Hence  for the coordinate ${\bf r}^{\prime\alpha}_\nu$  is related to ${\bf r}^\alpha$ as
\be
\label{tframes}
{\bf r}_{\nu}^\alpha={\bf r}^\alpha+{\bf r}^{\prime\alpha}_\nu.~~
\ee
Using (\ref{tframes}) in (\ref{cm-defn}), it follows that  
\be
\label{primecon}
\sum_{\nu=1}^s m^\alpha_\nu {\bf r}^{\prime\alpha}_{\nu}=0~~,
\ee
for the body-fixed coordinates.
The kinetic energy $ e_K^\alpha$  of the $\alpha$-th molecule is obtained as
\bea 
e_K^\alpha&=&\frac{1}{2}\sum_{\nu}m^\alpha_\nu(\dot{\bf
    r}^\alpha +\dot{\bf r}_\nu^{\prime\alpha}). (\dot{\bf r}^\alpha +\dot{\bf
    r}_\nu^{\prime\alpha})  \nonumber \\
&=& \frac{1}{2}\sum_\nu m^\alpha_\nu(\dot{\bf r}^\alpha+
\bm{\omega}^\alpha\times{\bf r}_\nu^{\prime\alpha}).(\dot{\bf r}^\alpha+
\bm{\omega}^\alpha\times{\bf r}_\nu^{\prime\alpha}) \nonumber \\
&=& \label{alg-e}
\frac{{\bf p}^\alpha.{\bf p}^\alpha}{2m^\alpha} +
\bm{\omega}^\alpha\cdot\frac{\rttensor{\bm{\kappa}}_\alpha}{2}
\cdot\bm{\omega}^\alpha~~.
\eea
The cross terms in the above expression on the right-hand side of Eq.
(\ref{alg-e}) vanishes due to (\ref{primecon}). 
In writing Eq. (\ref{alg-e}) we assume that 
$\rttensor{\bm{\kappa}}_\alpha$ is the moment of inertia tensor of rank two for the $\alpha$-th molecule, considered in a frame with the origin at its centre of mass. 
The total energy $\hat{e}^\alpha_K$ is obtained as
$\hat{e}^\alpha_K={e}^\alpha_\mathrm{T}+{e}^\alpha_\mathrm{R}$, 
where superscripts T and R denote the translational and
rotational part of the Kinetic energy of the $\alpha$-th
molecule, respectively denoted as,
\bea \label{e-tran} \hat{e}^\alpha_\mathrm{T} &=&
\frac{{\bf p}^\alpha.{\bf p}^\alpha}{2m^\alpha}\\
\label{e-rot}
\hat{e}^\alpha_\mathrm{R} &=& \frac{1}{2}\bm{\omega^\alpha}\cdot
\rttensor{\bm{\kappa}}_\alpha \cdot \bm{\omega^\alpha}~~~.\eea
If the molecules are distribution of the rigid set of elements with a cylindrical symmetry, then the molecule is treated like a pencil, and $\rttensor{\bm{\kappa}}_\alpha$ be chosen in
a diagonal form in terms of its principle components, such that
two of the diagonal elements of $\rttensor{\bm{\kappa}}_\alpha$ are
\{$\kappa_\alpha,\kappa_\alpha,\kappa_\alpha^\prime$\}. In case the distribution
is isotropic the $\rttensor{\bm{\kappa}}_\alpha$ is of  the form \{$\kappa_\alpha,\kappa_\alpha,\kappa_\alpha$\}. 
Since all the molecules are identical, we take $\kappa_\alpha=\kappa_0$ for all $\alpha=1,...N$.
For a rigid set of points $\{ {\bf r}^{\prime\alpha}_\nu\}$ for $\nu=1,...s$ comprising of the molecule $\alpha$, the angular speed is independent
of the origin. The angular momentum of the $\alpha$-th molecule then is obtained as
\be \label{angmom} \bm{\ell}^\alpha = \rttensor{\bm{\kappa}}_\alpha\bm{\omega^\alpha}~~. \ee
Therefore $\{{\bf v}^\alpha, \bm{\omega}^\alpha\}$ (for $\alpha=1,...,N$) are the linear and angular velocities of the $\alpha$ molecule. The angular velocity $\bm{\omega}^\alpha$ is measured in terms of the precision of the fixed length vector ${\bf u}$ around an axis in the direction of $\bm{\omega}^\alpha$.
This angular momentum $\bm{\ell}^\alpha$ is due to the rotational 
motion of the director ${\bf u}$. The $i$-th component of the angular momentum of the molecule $\alpha$ (=$1$, ...,$N$) is obtained as, $\ell^\alpha_i=\rttensor{\kappa}_{ij}\omega^\alpha_j$.  We assume that all the molecules are identical (hence $\rttensor{\kappa}_{ij}$ is independent of $\alpha$) and are cylindrically symmetric. So $\rttensor{\kappa}_{ij}$ has the form $\kappa^0_i\delta_{ij}$ where
$i$ denotes the principal axes. $\kappa^0_i$ is the corresponding principle moment of inertia. For an isotropic molecule $\kappa^0_i=\kappa_0$ and for a cylindrically symmetric one $\kappa^0_i{\equiv}\{\kappa_0,\kappa_0,\kappa_0^\prime\}$.

\subsection{Translational and Rotational motion}

The rotational dynamics are represented in the components of a fixed length vector ${\bf u}^\alpha$, located at the molecule's centre of mass $\alpha$ ( for $\alpha=1,...N$). The precession of ${\bf u}^\alpha$ in the body fixed frame signifies the rotational motion of the molecule $\alpha$. In three dimensions,  ${\bf u}^\alpha$ is described in terms of the two associated angles. The corresponding angular velocity $\bm{\omega}^\alpha(t)$ is related to rate of change of ${\bf u}^\alpha$ as,
\be \label{udot-e}
\frac{d{\bf u}^\alpha}{dt}=\bm{\omega}^\alpha{\times}{\bf u}^\alpha.~~\ee
%
%
%
The torque $\bm{\tau}^\alpha$ on the $\alpha$-th molecule at the centre of mass is obtained in terms of the director vector ${\bf u}^\alpha$ whose precession defines the rotational motion of the molecule.

\subsection{The Equations of motion}

We now consider the equations for the time evolution of the microscopic variables for the translational and rotational degrees of freedom in the fluid. The force equation for the Brownian motion of the $\alpha$-th molecule  ($\alpha=1,...N$) performing translational motion is written as \cite{debye,McConnel}:
\be \label{m-lin} \frac{d{\bf p}^\alpha}{dt} = -{\gamma}^{-1}_0 \frac{d{\bf r}^\alpha}{dt} + \bm{f}^\alpha+\bm{\eta}^\alpha~~.\ee
The frictional force is on the right-hand side of Eqn. (\ref{m-lin}) is proportional to the speed of the particle ${\bf v}^\alpha(=d{\bf r}^\alpha/dt$) with the proportionality constant $\gamma_0^{-1}$.
The force on particle $\alpha$ due to its interaction with other molecules (see Eq. (\ref{def-for}) below) is $\bm{f}^\alpha$. The stochastic part of the force on the $\alpha$-th particle in the fluid is given by $\bm{\eta}^\alpha$. The correlation of the noise is obtained as,
\be \label{for-cor}
<\eta_i^\alpha(t)\eta_j^\nu(t')>={2mT}{\gamma_0}^{-1}
\delta_{\alpha\nu}\delta_{ij}\delta(t-t'),~~\ee
where average momentum correlation of the particles is $<p_i^\alpha(t)p_j^\alpha(t)>=mT\delta_{ij}$.
The torque equations for the rotational Brownian motion are obtained as,
\be
\label{m-ang} \frac{d\bm{\ell}^\alpha}{dt} =- {\gamma}^{-1}_u \bm{\omega}^\alpha+ \bm{\tau}^\alpha+\bm{\zeta}^\alpha~~. \ee
The frictional component of the torque is proportional to the angular velocity $\bm{\omega}^\alpha$ of the $\alpha$-th particle with proportionality constant $\gamma_u^{-1}$. The
torque $\bm{\tau}^\alpha$  on the molecule $\alpha$ of the fluid is obtained from the intra-molecular interactions as defined in Eq. (\ref{def-tor}).
The fast part of the torque,  $\bm{\zeta}^\alpha$ correlates similar to $\bm{\eta}^\alpha$.  
\be \label{tor-cor}
<\zeta_i^\alpha(t)\zeta_j^\nu(t')>={2\kappa_0{T}}{\gamma_u}^{-1}
\delta_{\alpha\nu}\delta_{ij}\delta(t-t')~~. \ee
Eq. (\ref{m-lin})-(\ref{tor-cor}) describes the rotational and translational Brownian motions of the molecules in the fluid.

In the following, we present the dynamics analysis at two levels.
This description involving both the configurational-variables $\{{\bf r}^\alpha,{\bf u}^\alpha\}$ and the corresponding momentum-variables $\{{\bf p}^\alpha,\bm{\ell}^\alpha\}$, will be referred to as the Fokker-Planck dynamics. A reduced level description in terms of only the configuration variables 
$\{{\bf r}^\alpha,{\bf u}^\alpha\}$ will be referred to as Smoluchowski dynamics.
In the latter case, the momentum variables are eliminated from the description with the assumption of overdamped motion. Therefore, for the Smoluchowski dynamics, the inertial terms are on the right-hand side of the respective Eqns. (\ref{m-lin})-(\ref{m-ang}) are ignored, assuming that the corresponding momenta relax fast, and we obtain respectively,
\bea \label{ma-lin}  -{\gamma}^{-1}_0 \frac{d{\bf r}^\alpha}{dt} &=& \bm{f}^\alpha+\bm{\eta}^\alpha,~~\\
\label{ma-ang} - {\gamma}^{-1}_u \bm{\omega}^\alpha &=& \bm{\tau}^\alpha+\bm{\zeta}^\alpha~~.
\eea
Taking the cross product of Eqn. (\ref{ma-ang}) with vector ${\bf u}^\alpha$ and using Eqn. (\ref{udot-e}), we obtain from the approximate Eqn. (\ref{ma-ang}) the equation of motion for ${\bf u}^\alpha$.
\be 
\label{e0-u} \gamma^{-1}_u \frac{d{\bf u}^\alpha}{dt} =
\bm{\tau}^\alpha\times{\bf u}^\alpha +\bm{\zeta}^\alpha\times{\bf
    u}^\alpha ~~. \ee
The respective second terms in the right-hand sides of (\ref{ma-lin}) and Eqs. (\ref{e0-u}) signify the stochastic or fast parts of the corresponding dynamics. The noise is simple for the linear variable ${\bf r}$, while it is multiplicative for the orientational variable ${\bf u}$. The $i${-}th component of the random force in Eq. (\ref{e0-u}) is $\eps_{ijk}{\zeta^\alpha_j}{u^\alpha_k}$, and the presence of components of ${\bf u}^\alpha$ makes the noise multiplicative.

The equation of motion (\ref{e0-u}) for ${\bf u}^\alpha$ 
gives rise to a stochastic process\cite{uhlenbeck} described with a difference equation for ${\bf u}^\alpha(t)$ (over the time interval $t$ to $t{+}\Delta$) involving slow and fast parts. Since the noise in the Langevin equation 
(\ref{e0-u}) is multiplicative, the stochastic part of the corresponding difference equation has to be interpreted either with the I\^{t}o scheme\cite{ito} or the Stratonovich scheme \cite{staton}  for a complete description of the dynamics \cite{stoc-calculus}.
In Appendix \ref{app1}, we demonstrate that the difference equation corresponding to Eq. (\ref{e0-u}) with I\^{t}o interpretation of the noise is obtained using integrated noise $\bar{\bm{\zeta}^\alpha}$, defined in Eq. (\ref{av-noise}), as 
\be
\label{e0-ito} 
\gamma^{-1}_u \Big \{{\bf u}^\alpha(t{+}\Delta){-}{\bf u}^\alpha(t)\Big \}=
\int_t^{t+\Delta}d\bar{t} \{\bm{\tau}^\alpha(\bar{t}){\times}{\bf u}^\alpha(\bar{t})\} +\Delta^{\frac{1}{2}}\bar{\bm{\zeta}}^\alpha(t)\times{\bf
    u}^\alpha(t)~.   
\ee
However, with the Stratonovich interpretation of the multiplicative noise in Eq. (\ref{e0-u}), we show in Appendix \ref{app1} that the slow part, {\em, i.e.}, the first term in the right-hand side of  Eq. (\ref{e0-ito})) has an extra contribution with the replacement,
\be \label{e0-sta} 
\left \{ \mathbf{\tau}^\alpha{\times}{\bf u}^\alpha  \right \}{\rightarrow}
\left \{ \mathbf{\tau}^\alpha{\times}{\bf u}^\alpha{-}2T{\bf u}^\alpha \right \}
\ee
On the other hand for the I\"{t}o interpretation of the multiplicative noise in Eq. (\ref{e0-u}) the term $-2T{\bf u}^\alpha$ in the right-hand side of Eqn. (\ref{e0-sta}) is not required.

\subsection{Collective  densities: microscopic definitions}

The property of a many-particle system is often described in terms of a set
of local densities characteristic of the system. The collective density in terms of which the dynamics are formulated in this case is the density $\hat{\rho}({\bf r},{\bf u},t)$. At the microscopic level description, it
is denoted with a hat and is defined in terms of the configuration variables
as follows.
\be \label{def-rho} \hat{\rho}({\bf r},{\bf u},t) =
\sum_{\alpha=1}^N \delta({\bf r}-{\bf r}^\alpha(t)) \delta({\bf
    u}-{\bf u}^\alpha(t)) {\equiv} \sum_{\alpha=1}^N
\delta({\bf x}-{\bf x}^\alpha(t)) ~~. \ee
For the set of variables ${\bf x}^\alpha=\{{\bf r}^\alpha,{\bf
    u}^\alpha\}$, the corresponding external labels are $\{{\bf  r},{\bf u}\}$
which we will denote as ${\bf x}$. 
The product of the delta functions on the
right-hand side of Eqn. (\ref{def-rho}) as $\delta({\bf x}-{\bf
    x}^\alpha(t))$. 
The collective densities in the Fokker-Planck dynamics involve an extended set including the linear and angular momentum densities, respectively denoted as ${\bf g}({\bf x},t)$ and $\bm{\ell}({\bf x},t)$. These two momentum densities are  respectively defined as
\bea \label{def-g} \hat{\bf g}({\bf x},t) &=& \sum_{\alpha=1}^N
{\bf p}^\alpha(t) \delta({\bf x}-{\bf x}^\alpha(t))~, \\
\label{def-l} \hat{\bm{\ell}}({\bf x},t)   &=& \sum_{\alpha} 
\bm{\ell}^\alpha(t) \delta({\bf x}-{\bf x}^\alpha(t)). \eea
The total energy density is also obtained in an exactly similar way as,
a sum of translational and rotational parts:
\be
\label{cons-edef} \hat{e} ({\bf x},t) =
\sum_\alpha\{{e}_\alpha^\mathrm{T}+{e}_\alpha^\mathrm{R}\} \delta
({\bf x} - {\bf x}^\alpha(t)){\equiv} \hat{e}_\mathrm{T}({\bf
    x},t)+\hat{e}_\mathrm{R}({\bf x},t)~~,
\ee
where $\hat{e}^\alpha_\mathrm{T}$ and $\hat{e}^\alpha_\mathrm{R}$ are respectively the translational and rotational parts of the energy of the 
particle $\alpha$ defined in Eqns. (\ref{e-tran}) and (\ref{e-rot}). 
In the following, we formulate the dynamics of a many-particle system in terms of a suitable set of collective variables defined above. First, we will consider the reduced description with the configuration variables (Brownian Dynamics) and next with the so-called Fokker-Planck dynamics with the extended configuration and momentum variables.

\section{Dynamics of the collective variables:}

For the case of Smoluchowsky dynamics, the equation of motion is obtained for the collective density $\hat{\rho}({\bf x},t)$. For the Fokker-Planck dynamics, the corresponding collective variables are extended to the set $\{\hat{\rho}(\bfx,t),\hat{\bf g}(\bfx,t),\hat{\bm{\ell}}(\bfx,t)\}$.

\subsection{The Brownian Dynamics}
For the Brownian dynamics, the time evolution of the system is formulated in terms of only the configuration variables, {\em i.e.}, the position and angular coordinates ${\bf x}^\alpha$ ${\equiv}$ $\{{\bf r}^\alpha,{\bf u}^\alpha\}$, and is similar to the Dean-Kawasaki dynamics used for only translational degrees of freedom.
To obtain an equation for the density $\hat{\rho}({\bf x},t)$ defined above, we follow the Dean's method starting from
the definition (\ref{def-rho}) and the microscopic equation of motions (\ref{ma-lin}) and (\ref{e0-sta}). Here, we use the
following chain rule of stochastic differential equations I\^{t}o
calculus \cite{oksendal,stoc-calculus}. We consider the dynamics  for set of stochastic variables $x^\alpha_i(t)$ ($\alpha=1,..N$) as:
\be \label{itoc1} {\dot{x}}_i^\alpha=h_i^\alpha +\sum_\nu g_{ij}^{\alpha\nu}\xi_j^\nu~~,
\ee
where the correlation of the white noise $\xi_i$ ( with I\^{t}o interpretation ) is defined as,
\be \label{itoc2}
<\xi_i^\alpha(t)\xi_j^\nu(t')>=\delta_{\alpha\nu}\delta_{ij}\delta(t-t')~~. \ee
In this case the matrix $g_{ij}^{\alpha\delta}$ is diagonal in
particle indices $\alpha$, and $\nu$ since
the corresponding microscopic equations of motion (\ref{ma-lin}) and (\ref{e0-sta})  do not involve coupling between various particles in their 
random components respectively denoted
by $\eta^\alpha$ and $\zeta^\alpha$. Thus $g_{\alpha\nu}$ has a
block diagonal form
\be \label{gb-diag} g_{ij}^{\alpha\nu}=\left [
\sqrt{2{\gamma_0}T}\delta_{ij},
\sqrt{2{\gamma_u}T}\eps_{ijk}u_k^{\alpha} \right ]
\delta_{\alpha\nu}, \ee
$u_i^{\alpha}$ being the $i$-th direction Cartesian component of
vector ${\bf u}^\alpha$, and the Levi-Civita symbols $\eps_{ijk}$
are defined by properties like $\eps_{ijk}=-\eps_{jik}=1$, when
$i,j,k$ are in cyclic order, and $\eps_{iik}=0$.
The I\^{t}o chain rule obtains the stochastic differential
equation for the variable $y(\{x_i\})$  in the form
\begin{equation}
    \label{Ito-cal3} \dot{y} = \sum_\alpha \dot{x}_i^\alpha
    \frac{\partial{y}}{\partial{x_i^\alpha}} +
    \sum_{\alpha,\nu,\delta}
    \frac{1}{2}\frac{\partial^2{y}}{\partial{x_i^\alpha}\partial{x_j^\nu}}
    g_{ik}^{\alpha\delta}g_{jk}^{\nu\delta}~~.
\end{equation}
Eq. (\ref{Ito-cal3}) follows directly by taking
the deviation of the function $y(x^\alpha)$ in terms of the
corresponding variation in the $x^\alpha$'s. We apply the
chain rule  (\ref{Ito-cal3}) for the stochastic 
variables ${\bf x}^\alpha(t)$ for $\alpha=1,..N$ with the
respective equations of motion being given by  (\ref{ma-lin}) 
and (\ref{e0-sta}).  We define in this case the
function $y$ as, $y(\{{\bf x}^\alpha\}) \equiv \hat{\rho}({\bf
    x},t)$, where $\hat{\rho}$ is as defined in Eqn.  (\ref{def-rho}).
Note that the coordinates ${\bf x}$ acts as a label on
$\hat{\rho}$ which is also a function of $\{{\bf x}^\alpha\}$.
The corresponding stochastic
differential equation for the density variable $\hat{\rho}({\bf
    x},t)$ is obtained as,
\be \label{BD-re1}\frac{\pd}{\pd t}\hat{\rho}({\bf x},t)
{=}\sum_{\alpha=1}^N\Bigg \{ \dot{\bf r}_\alpha\cdot \nabla_{{\bf r}^\alpha}+ 
\dot{\bf u}^\alpha \cdot \nabla_{{\bf u}^\alpha} \Bigg \}\delta({\bf x}-{\bf  x}^\alpha(t)) 
+{\cal I}_c
\ee
The quantity ${\cal I}_c$ on the right-hand side of Eq. (\ref{BD-re1}) arises from the double derivative term (second term on the right-hand side)  of the I\^{t}o formula (\ref{Ito-cal3}) and is obtained as,
\be 
\label{id-term}
{\cal I}_c=\sum_{\alpha=1}^N 
\left [ {\gamma_0}T\nabla^2_{{\bf  r}^\alpha}+{\gamma_u}T\epsilon_{ijk}\epsilon_{i'jk'}
{u_k^\alpha}{u_{k'}^\alpha}\nabla_{{\bf u}^\alpha}^{i}
\nabla_{{\bf u}^\alpha}^{i'}\delta({\bf x}-{\bf  x}^\alpha(t)) \right ]
\ee
For the Levi-Civita symbols $\epsilon_{ijk}$ etc. the following identity holds:
\be \label{LC-r1}
\epsilon_{ijk}\eps_{i'jk'}{u_k^\alpha}{u_{k'}^\alpha}\bm{\nabla}_{{\bf
        u}^\alpha}^{i} \bm{\nabla}_{{\bf u}^\alpha}^{i'} =
\epsilon_{ijk}\eps_{ijk}{u_k^\alpha}{u_k^\alpha}\bm{\nabla}_{{\bf
        u}^\alpha}^i \bm{\nabla}_{{\bf u}^\alpha}^i
+\eps_{ijk}\eps_{kji}{u_k^\alpha}{u_i^\alpha}\bm{\nabla}_{{\bf
        u}^\alpha}^i \bm{\nabla}_{{\bf u}^\alpha}^k~~. \ee
Furthermore, we note that the derivative operators like
$\nabla_{{\bf r}^\alpha}$ or $\bm{\nabla}_{{\bf u}^\alpha}$ acting on
the delta functions in the right-hand side of Eqn. (\ref{BD-re1})
can be replaced as derivative operators with the respective
external labels {\em, i.e.} as $-\nabla_{\bf r}$ and $-\bm{\nabla}_{\bf u}$.
Here, we have used the notation in which both $\bm{\nabla}_{\bf r}$ and $\bm{\nabla}_{\bf u}$ have the same dimensions.
\be
\label{LC-r2} \sum_{\alpha=1}^N
\{\epsilon_{ijk}\eps_{i'jk'}{u_k^\alpha}{u_{k'}^\alpha}
\bm{\nabla}_{{\bf u}^\alpha}^{i} \bm{\nabla}_{{\bf
        u}^\alpha}^{i'}\delta({\bf x}-{\bf x}^\alpha(t))\}={\nabla_{\bf u}^i}{\nabla_{\bf
        u}^j}\{\Pi_{ij}\hat{\rho}({\bf x},t)\} ~~.\ee
The quantity  ${\Pi}_{ij}({\bf u})$ in the right hand side of Eq. (\ref{LC-r2}) represent the elements of the matrix $\bm{\Pi}$, and is defined as
\be \label{Pi-def}
\Pi_{ij}(u)=u^2\delta_{ij}-u_iu_j~~.
\ee
This substitution obtains for the corresponding derivative,
\be
\label{pi-math}
\nabla_{\bf u}^j\Pi_{ij}=2u_i-\delta_{ij}u_j-u_i\delta_{jj}=-2u_i
\ee
The equation of motion for the density $\hat{\rho}({\bf x},t)$is obtained
as
\bea \label{BD-stat}\frac{\pd\hat{\rho}}{\pd t}&=&
{\gamma_0}T\nabla_{\bf r}^2\hat{\rho}{+}\gamma_u
{T}{\nabla_{\bf u}^i}{\nabla_{\bf u}^j} \Big \{\Pi^{ij} (u) \hat{\rho}\Big \} 
{-} \Big \{ {\gamma_0}\nabla_{\bf r}^i \hat{\bm{f}}_i {+} \gamma_u\nabla_{\bf u}^i {\left [ \widehat{\bm{\tau}} \times{\bf u}
    -2T{\bf u}  \right ]}_i \Big \} \hat{\rho} \nonumber  \\
&-&\left [ \gamma_0\nabla_{\bf r} {\cdot} \hat{\bm{\eta}}{+}
\gamma_u\nabla_{\bf u}\cdot \{
\hat{\bm{\zeta}}\times {\bf u} \} \right ]~~.  \eea
The force-density, $\hat{\bm{f}}({\bf x},t)$ and torque-density $\widehat{\bm{\tau}}({\bf x},t)$ in the right hand side of Eqn. (\ref{BD-stat}) 
are respectively defined as,
\bea \label{f-term} \hat{\bm{f}}_i({\bf x},t)\hat{\rho}({\bf x},t)
&=&\sum_{\alpha=1}^N{\bm{f}}_i^\alpha \delta({\bf x}-{\bf
    x}^\alpha) \\
\label{t-term} 
\widehat{\bm{\tau}}_i({\bf x},t) \hat{\rho}({\bf x},t)
&=& \sum_{\alpha=1}^N {\bm{\tau}}_i^\alpha
\delta({\bf x}-{\bf x}^\alpha)~~. \eea
Note that in writing, the third term is on the right-hand side of Eqn. (\ref{BD-stat}) we have followed the Stratonovich interpretation of the multiplicative noise (See Eq. (\ref{e0-sta}), which is the equation of motion (\ref{e0-u}) for the microscopic dynamics of ${\bf u}$.
With the I\^{t}o interpretation of the noise in Eq. (\ref{e0-u}) for ${\bf u}$ we have the 
$-2T{\bf u}$ absent and Eq. 32 reduces to:
\bea \label{BD-ito}\frac{\pd\hat{\rho}}{\pd t}&=&
{\gamma_0}T\nabla_{\bf r}^2\hat{\rho}{+}\gamma_u
{T}{\nabla_{\bf u}^i}{\nabla_{\bf u}^j} \Big \{\Pi^{ij} (u) \hat{\rho}\Big \} 
{-} \Big \{ {\gamma_0}\nabla_{\bf r}^i \hat{\bm{f}}_i {+} \gamma_u\nabla_{\bf u}^i {\left [ \widehat{\bm{\tau}} \times{\bf u}\right ]}_i \Big \} \hat{\rho} \nonumber  \\
&-&\left [ \gamma_0\nabla_{\bf r} {\cdot} \hat{\bm{\eta}}{+}
\gamma_u\nabla_{\bf u}\cdot \{
\hat{\bm{\zeta}}\times {\bf u} \} \right ]~~.  \eea
The last two terms in the right-hand side of Eqn. (\ref{BD-stat}) (or Eq. (\ref{BD-ito}) )
are divergences (respectively with respect to components of ${\bf r}$ and ${\bf u}$) of two collective noise-fields which are defined as 
\bea \label{noise-r} \hat{\bm{\eta}}({\bf x},t)
&=&\sum_{\alpha=1}^N{\bm{\eta}^\alpha} \delta({\bf x}-{\bf x}^\alpha) \\
\label{noise-z}\hat{\bm{\zeta}}({\bf x},t) 
&=& \sum_{\alpha=1}^N \bm{\zeta}^\alpha
\delta({\bf x}-{\bf  x}^\alpha)~~. \eea
A single stochastic force field $\theta(\bfx,t)$ defined combining $\{\hat{\eta}({\bf x},t),\hat{\zeta}({\bf x},t)\}$ as:
\be \label{tot-noise} \hat{\theta}({\bf x},t)=-{\gamma_0}\nabla_{\bf r} \cdot
\hat{\bm{\eta}}({\bf x},t) - {\gamma_u}\nabla_{\bf u}
{\cdot}\hat{\bm{\zeta}}^{\bf u}({\bf x},t)~~,
\ee
where we have defined $\hat{\bm{\zeta}}^{\bf u}{\equiv}\hat{\bm{\zeta}}{\times}{\bf u}$.
Using the microscopic relations (\ref{for-cor}) and (\ref{tor-cor}) the noise correlations are obtained as,
\bea \label{ncor-rr} \langle \hat{\eta}_i({\bf x} ,t)
\hat{\eta}_j({\bf x}^\prime,t^\prime) \rangle &=&
2mT{\gamma_0^{-1}}\delta_{ij}\hat{\rho}({\bf
    x},t)\delta({\bf x}-{\bf x}')\delta(t-t')\\
\langle \hat{\zeta}^{\bf u}_i({\bf x} ,t) \hat{\zeta}^{\bf u}_j({\bf x}^\prime,t^\prime)
\rangle &=&
\sum_{\alpha=1}^N 2T{\gamma_u^{-1}}\eps_{ij'k}u^\alpha_{k}
\delta({\bf x}-{\bf x}^\alpha)\eps_{jj'k'}u_{k'}^{\alpha}
\delta({\bf x}^\prime-{\bf x}^\alpha) \delta(t-t')\nonumber\\
\label{ncor-uu} &=& 2T{\gamma_u^{-1}}\Pi_{ij}(u) \hat{\rho}({\bf
    x},t) \delta({\bf x}-{\bf x}')\delta(t-t') \eea
The cross-correlations for the noises are zero, {\em i.e.},
$\langle \hat{\eta}_i({\bf x} ,t) \hat{\zeta}_j({\bf x}^\prime,t) \rangle = 0$.
Correlation of the noise $\theta({\bfx},t)$  defined in Eq.  (\ref{tot-noise}) is obtained using the relations (\ref{ncor-rr})-(\ref{ncor-uu}) as,
\be
\label{ncor-tt-sta}
\langle \hat{\theta}({\bf x} ,t) \hat{\theta}({\bf x}^\prime,t) \rangle =
2T\nabla_{\bf x}^i{\cal D}_{ij}\hat{\rho}\nabla_{{\bf x}^\prime}^j
\delta({\bf x}-{\bf x}')\delta(t-t') ~~.
\ee
${\cal D}_{ij}$ in right hand side of Eq. (\ref{ncor-tt-sta}) denotes $2d\times{2d}$ transport matrix for the $d$ dimensional fluid, , and is obtained in the block diagonal form,
\be \label{diffu-matrix} {\cal D} = \left [ 
\begin{array}{c|c} 
    \gamma_0{\bf I}   & \bm{0} \\ 
    \hline 
    \bm{0} & \gamma_u\bm{\Pi}(u) 
\end{array} 
\right ]
\ee
where ${\bf I}$ is  $d{\times}d$ identity matrix, and $\bm{\Pi}$ is defined in (\ref{Pi-def}).

For the collective density, the equation of motion (\ref{BD-stat}) corresponding to the Stratonovich interpretation of the noise in the ${\bf u}$-equation is obtained in a compact form in terms of a current ${\bf J}$ and noise $\hat{\theta}$:
\be
\label{BD-comp}
\frac{\pd\hat{\rho}}{\pd t} = \bm{\nabla}_{\bf x}{\cdot}
\hat{\bf J}{+}\hat{\theta}~~.
\ee
The regular or slow part on the right-hand side of Eq. (\ref{BD-comp}) is obtained as the divergence of the current ${\bf J}$:
\bea
\label{j-total}
\bm{\nabla}_{\bf x}{\cdot}\hat{\bf J}(\bfx,t)&{=}&  
\gamma_0T{\nabla}_{\bf r}^i
{\nabla}_{\bf r}^i\hat{\rho}(\bfx,t){+}\gamma_uT{\nabla}^i_{\bf u}
({\Pi}_{ij} {\nabla}^j_{\bf u}\hat{\rho}(\bfx,t)) \nonumber \\
&{-}&\gamma_0\bm{\nabla}_{\bf r}{\cdot}\hat{(\bm{f}}(\bfx,t)\hat{\rho}){-}\gamma_u\bm{\nabla}_{\bf u}{\cdot}(\{ \hat{\bm{\tau}}(\bfx,t){\times}{\bf u}\}\hat{\rho}(\bfx,t)) 
\eea
We have used the relation (\ref{pi-math}) in obtaining the right-hand side of Eq. (\ref{j-total}). Eqn. (\ref{BD-ito}) corresponding to the I\^{t}o interpretation of the multiplicative noise in the ${\bf u}$-equation (\ref{e0-u}), is obtained in the same form (\ref{BD-comp})) with the current ${\bf J}$ replaced by ${\bf J}^\prime$ which is obtained as,
\bea
\label{j-total-ito}
\bm{\nabla}_{\bf x}{\cdot}\hat{\bf J}^\prime(\bfx,t)&{=}&  
\gamma_0T{\nabla}_{\bf r}^i
{\nabla}_{\bf r}^i\hat{\rho}(\bfx,t){+}\gamma_uT{\nabla}^i_{\bf u}{\nabla}^j_{\bf u}
\{{\Pi}_{ij}\hat{\rho}(\bfx,t))\} \nonumber \\
&{-}&\gamma_0\bm{\nabla}_{\bf r}{\cdot}\hat{(\bm{f}}(\bfx,t)\hat{\rho}){-}\gamma_u\bm{\nabla}_{\bf u}{\cdot}(\{ \hat{\bm{\tau}}(\bfx,t){\times}{\bf u}\}\hat{\rho}(\bfx,t)) 
\eea
The right hand sides of (\ref{j-total-ito}) has an extra contribution of $\gamma_u{T}\nabla_{\bf u}^i\{\nabla_{\bf u}^j \Pi_{ij}\}\hat{\rho}(\bfx,t)$ compared to that of
(\ref{j-total}).
The correlation of noise $\hat{\theta}(\bfx,t)$ is given by the equation (\ref{ncor-tt-sta}) and is multiplicative due to the factor $\hat{\rho}(\bfx,t)$ in the weight of the correlation.

\subsection{Hamiltonian Systems}

The stochastic dynamics of the collective density $\hat{\rho}(\bfx,t)$ is obtained from the equations of motions (\ref{m-lin}) and (\ref{m-ang}), respectively involving driving forces and torques. The corresponding random parts in the translational and rotations Brownian motion are respectively denoted as  $\bm{\eta}$ and $\bm{\zeta}$, while the frictional forces in the are proportional to the linear and angular velocities of the particles. We now analyze the right-hand side of Eq. (\ref{BD-stat}) as obtained in Eqn. (\ref{BD-comp}), and in Eqn. (\ref{j-total}) corresponding to Statonvitch's interpretation of the noise in Eq. (\ref{e0-u}). The case for the I\^{t}o interpretation (with Eq. (\ref{BD-comp}) and (\ref{j-total-ito})) is considered next. 
We write the current {\bf J} as sum of two parts ${\bf J}{=}{\bf J}_\mathrm{id}+{\bf J}_\mathrm{in}$. The part ${\bf J}_\mathrm{id}$ involves the first two terms on the right-hand side of (\ref{j-total}), and its divergence is expressed in terms of functional derivative with respect to density $\hat{\rho}(\bfx)$ of a function $\widehat{F_{\rm id}}[\hat{\rho}]$:
\bea
\label{fder1}\bm{\nabla}_{\bf x}{\cdot}\hat{\bf J}_\mathrm{id}[\hat{\rho}]&=&
\Big [ {\gamma_0}T\nabla^2_{\bf r} {+} {\gamma_u}T\nabla^i_{\bf u}
{\Pi}_{ij}(u)\nabla^j_{\bf u}\Big ] \hat{\rho}({\bf x},t)\}
\nonumber \\
&=&  \gamma_0T\nabla_{\bf r}\cdot \Big \{\hat{\rho}\nabla_{\bf r}
\frac{\delta{\widehat{F}_{\rm id}}}{\delta\hat{\rho}}\Big \}{+}{\gamma_u}T
\nabla^i_{\bf u}\cdot\Big \{\Pi_{ij}(u)\hat{\rho}\nabla_{\bf u} ^j
\frac{\delta{\widehat{F}_{\rm id}}}{\delta\hat{\rho}} \Big \}~~.
\eea
where 
\be
\label{fen-ide}
\widehat{F}_{\rm id}[\hat{\rho}]{=} \int d {\bf x} \hat{\rho}({\bf x})
[\ln\hat{\rho}({\bf x})-1]~.
\ee
The last two terms on the right-hand side of (\ref{j-total}) are expressed in terms of a functional derivative with respect to density $\hat{\rho}(\bfx)$ of the functional $F_\mathrm{in}[\hat{\rho}]$ which is obtained in terms of the microscopic interactions among the particles. 
We assume that the force $\bm{f}$ and the torque $\bm{\tau}$ can be obtained in terms of the inter-particle interaction.

The Hamiltonian $H$\cite{LL-v1} for the many-particle system is obtained as $H=K+U$, a sum of kinetic and potential parts denoted by $K$ and $U$. 
The kinetic part $K$ is the sum of the kinetic energy of each element, $K=\sum_\alpha e_K^\alpha$, where
the kinetic energy of the $\alpha$-th molecule is obtained as $e_K^\alpha$.
The interaction energy $U$ is introduced in equation (\ref{pe1}).
The pairwise interaction part $U_0$, defined in Eqn. (\ref{ipote}) is obtained in terms of the collective density function $\hat{\rho}$ as
\be
\label{pote}
U_0{=}\frac{1}{2}
\int d {\bf x} \int d {\bf x}^\prime \Phi_0({\bf x}-{\bf x}^\prime)
\hat{\rho}({\bf x}) \hat{\rho}({\bf x}^\prime) ~.
\ee
The single particle contributions are expressed as $\Phi_1(\bfx^{\alpha})$, which is dependent on the position ${\bf x}^\alpha$ for the single molecule $\alpha$. 
\be \label{pe} U_1=\sum_{\alpha=1}^N \Phi_1({\bf x}^\alpha)=
\int d {\bf x} \Phi_1({\bf x})\hat{\rho}({\bf x}) ~. \ee
The force ${\bm{f}}^\alpha$ on the $\alpha$-th molecule is therefore obtained as,
\be
\label{def-for}
{\bm{f}}_{\bf r}^\alpha = -\Big ( \frac{\partial{U}}{\partial{\bf r}^\alpha}\Big )=
- \nabla_{{\bf r}^\alpha}
\left [ \sum_{\delta=1}^{N\ \prime}\Phi_0({\bf x}^\alpha-{\bf x}^\delta){+}\Phi_1({\bf x}^\alpha)\right ]~.
\ee
$\bm{f}^\alpha$  gives rise to the translational motion. 
Using the above expression $\bm{f}^\alpha$, we obtain from  Eqn. (\ref{f-term}) the following result,
\bea
\label{fder2}
&&\hat{\rho}({\bf x},t)\hat{\bm{f}}({\bf x},t)= 
-\sum_\alpha
\delta({\bfx}-{\bfx}_\alpha)
\nabla_{{\bf r}^\alpha}
\left [ \sum_{\delta=1}^{N\ \prime}\Phi_0({\bf x}^\alpha-{\bf x}^\delta){+}\Phi_1({\bf x}^\alpha)\right ]~ \nonumber \\
&=&-\sum_\alpha
\delta({\bfx}-{\bfx}_\alpha)
{\int}d{\bfx}_1
\left [ \sum_{\nu=1}^{N\ \prime}{\int}d{\bfx}_2\Phi_0({\bf x}_1-{\bf x}_2)\delta({\bfx}_2-{\bfx}_\nu){+}\Phi_1({\bf x}_1)\right ]\nabla_{{\bf r}^\alpha}
\delta({\bfx}_1-{\bfx}_\alpha)~ \nonumber \\
&=&\sum_\alpha
\delta({\bfx}-{\bfx}_\alpha)
{\int}d{\bfx}_1
\left [ {\int}d{\bfx}_2\Phi_0({\bf x}_1-{\bf x}_2)\hat{\rho}(\bfx_2){+}\Phi_1({\bf x}_1)\right ]\nabla_{{\bf r}_1}
\delta({\bfx}_1-{\bfx})~ \nonumber \\
\label{tor-lin}
&=&-\hat{\rho}({\bf x},t)\nabla_{\bf r}\Big [ \int d{\bf x}^\prime\Phi_0({\bf x}-{\bf x}^\prime)\hat{\rho}({\bf x}^\prime){+}\Phi_1(\bfx)\Big ]
\equiv-\hat{\rho}\nabla_{\bf r}\frac{\delta{\widehat{F}_{\rm in}}}{\delta\hat{\rho}}
\eea
where the functional $\widehat{F}_{\rm in}$ is  defined as,
\be
\label{fen-int}
{\beta}\widehat{F}_{\rm in}[\hat{\rho}]{=}\frac{1}{2}
\int d {\bf x} \int d {\bf x}^\prime \Phi_0({\bf x}{-}{\bf x}^\prime)
\hat{\rho}({\bf x}) \hat{\rho}({\bf x}^\prime){+}
\int d {\bf x} \Phi_1({\bf x})\hat{\rho}({\bf x}) 
\ee
For the angular dependence through the variable ${\bf u}$, we obtain the corresponding 
torque
%
\be \label{tor-alpha} 
{\bm{f}}_u^\alpha=-\frac{\pd{U}}{\pd{\bf u}^\alpha}{=} -\frac{\pd}
{\pd{\bf u}^\alpha}\left [ \sum_{\nu=1}^{N\ \prime} {\Phi_0({\bf x}^\alpha-{\bf x}^\nu)}
+{\Phi_1({\bf x}^\alpha)}\right ]~~. 
\ee
Hence, the torque around the centre of mass of the molecule $\alpha$ is,
\be
\label{def-tor}
\bm{\tau}^\alpha = {\bf u}^\alpha \times{\bm{f}}_u^\alpha {=}-{\bf u}^\alpha \times \frac{\pd}{\pd{\bf u}^\alpha}
\left [ \sum_{\nu=1}^{N\ \prime} {\Phi_0({\bf x}^\alpha-{\bf x}^\nu)}
+{\Phi_1({\bf x}^\alpha)}\right ]~~.
\ee
Using the definition (\ref{t-term}) for $\hat{\bm{\tau}}(\bfx,t)$ we obtain,
\bea
\label{t1-term} 
\widehat{\bm{\tau}}({\bf x},t) \hat{\rho}({\bf x},t)&=&-{\sum_{\alpha=1}^N}
\Bigg \{ {\bf u}^\alpha{\times}\frac{\pd}{\pd{\bf u}^\alpha}
\left ( \sum_{\nu=1}^{N\ \prime} {\Phi_0({\bf x}^\alpha-{\bf x}^\nu)}
+{\Phi_1({\bf x}^\alpha)}\right ) \Bigg \}
\delta({\bf x}-{\bf x}^\alpha)~~\nonumber \\
&=&  \Bigg \{ {\bf u}{\times}\frac{\pd}{\pd{\bf u}}
\left [ \sum_{\nu=1}^{N\ \prime} {\Phi_0({\bf x}-{\bf x}^\nu)}
+{\Phi_1({\bf x})}\right ]\Bigg \}
\sum_{\alpha=1}^N \delta({\bf x}-{\bf x}^\alpha)~ \nonumber \\
&=&  \Bigg \{{\bf u}{\times}\bm{\nabla}_{\bf u}
\left [ \frac{\delta{\widehat{F}_{\rm in}}}{\delta\hat{\rho}({\bf x},t)}\right ]\Bigg \}\hat{\rho}(\bfx,t)~ .
\eea
Therefore, the last term on the right-hand side of Eq, (\ref{j-total}) is obtained as 
\bea
\widehat{\bm{\tau}}({\bf x},t)\hat{\rho}({\bf x},t){\times}{\bf u}
&=&- \left \{ {\bf u}{\times}{\bf u}{\times}\bm{\nabla}_{\bf u}
\left ( \frac{\delta{\widehat{F}_{\rm in}}}{\delta\hat{\rho}({\bf x},t)}\right )
\right \}\hat{\rho}(\bfx,t)~ \nonumber \\
&=&\Bigg \{ (u_iu_j-u^2\delta_{ij})\nabla_{\bf u}^j
\left ( \frac{\delta{\widehat{F}_{\rm in}}}{\delta\hat{\rho}({\bf x},t)}\right )
\Bigg \} \hat{\rho}({\bf x},t) \nonumber \\
\label{tor-ex2}
&{\equiv}& -\hat{\rho}({\bf x},t)\Pi_{ij}(u)\nabla_{\bf u} ^j 
\frac{\delta{\widehat{F}_{\rm in}}}{\delta\hat{\rho}({\bf x},t)}~~.
\eea
Finally, using (\ref{fder1}), (\ref{tor-lin}) and (\ref{tor-ex2}), the equation (\ref{BD-comp}) for the time evolution $\hat{\rho}({\bf x},t)$ corresponding to Statonvitch interpretation of the multiplicative noise in the ${\bf u}$-equation (\ref{e0-u}) reduces to the form:
\be
\label{BD-re5}
\boxed{\frac{\pd}{\pd t}\hat{\rho}({\bf x},t)
    {=}\bm{\nabla}_{\bf x}{\cdot}{\bf J}{=}\nabla_{\bf x}^i {\cal D}_{ij}\hat{\rho}\nabla_{\bf x}^j
    \left (
    \frac{\delta\widehat{F}[\hat{\rho}]}{\delta\hat{\rho}
        ({\bf x},t)}\right )+\hat{\theta}({\bf x},t) }~~.
\ee
The stochastic field $\hat{\theta}(\bfx,t)$ is defined in Eqn. (\ref{tot-noise}) and
its correlation is given by Eq.  (\ref{ncor-tt-sta}).
The transport matrix ${\cal D}$ is as defined in Eqn. (\ref{diffu-matrix}), and the functional $F[\hat{\rho}]$ is expressed as a sum of an ideal gas part $F_{\rm id}[\hat{\rho}]$  and an interaction part $F_{\rm in}[\hat{\rho}]$, 
\be \label{fen-total} F[\hat{\rho}]{=}F_{\rm id}[\hat{\rho}] + F_{\rm in}[\hat{\rho}],
\ee
The functional $\widehat{F}$s depends on phase space coordinates involving $\delta$-functions. For the I\^{t}o interpretation of the noise, the Eqns. (\ref{BD-comp}) and (\ref{j-total-ito})) are considered and the resulting equation for $\rho({\bfx},t)$ is obtained in the same form as Eqn. (\ref{BD-re5}) with two different contributions 
\be
\label{BD-re5-ito}
\frac{\pd}{\pd t}\hat{\rho}({\bf x},t)
{=}\nabla_{\bf x}^i \widetilde{\cal D}_{ij}\hat{\rho}\nabla_{\bf x}^j
\left ( \frac{\delta\widehat{F}[\hat{\rho}]}{\delta\hat{\rho}({\bf x},t)}\right ){+}
\nabla_{\bf x}^i {\cal D}^0_{ij}\hat{\rho}\nabla_{\bf x}^j
\left (
\frac{\delta\widehat{F}^\prime[\hat{\rho}]}{\delta\hat{\rho}
    ({\bf x},t)}\right )    
+\hat{\theta}({\bf x},t) ~~.
\ee
We have now defined the elements of the transport matrix ${\cal D}$ in terms of a traceless part $\widetilde{\cal D}$, and the trace ${\cal D}^0$ respectively defined 
for $d$ spatial dimensions as follows:
\bea
\label{tr-def1}
\widetilde{\cal D}&=&{\cal D}-{\cal D}^0~.\\
\label{tr-def2}
{\cal D}^0&=&\frac{1}{d} \left [ 
\begin{array}{c|c} 
    \gamma_0{d}{\bf I}  & O \\ 
    \hline 
    O & {\gamma_uu^2}{(d{-}1)}{\bf I}
\end{array} 
\right ] 
\eea
The functional $\widehat{F}^\prime$ introduced on the right hand side of Eq. (\ref{BD-re5-ito})  is obtained as,
\be \label{fenh-ito} \widehat{F}^\prime{=} \widehat{F}{-}{d}{\int}{d}{\bf x} {\hat{\rho}({\bf x})}
{\ln {u}}.
\ee
To summarize the above results: 
For the I\^{t}o interpretation of the multiplicative noise in the  ${\bf u}$-equation, the relation (\ref{fenh-ito})  between the two quantities $\widehat{F}$ and $\widehat{F}^\prime$ holds, and the equation of motion for the collective density $\hat{\rho}(\bfx,t)$  is given by (\ref{BD-re5-ito}). For the Stratonovich interpretation of the multiplicative noise in the ${\bf u}$-equation, the two quantities $\widehat{F}$ and $\widehat{F}^\prime$ are the same, and we recover the equation of motion  (\ref{BD-re5}) in agreement with recent works \cite{cugli}.

\subsection{Fokker-Planck (FP) dynamics }

We now formulate closed equations in the under-damped case, which will be
referred to as the Fokker-Planck (FP) dynamics,
in which a larger set of variables $\{{\bf x},{\bf p}\}$ , involving
both position and momentum densities are included.
Here, ${\bf p}^\alpha=m\dot{\bf r}^\alpha$ is the linear momentum and  $\bm{\ell}^\alpha(t)=\rttensor{\bm{\kappa}}_\alpha\bm{\omega}^\alpha$  is the angular momentum of the particle $\alpha$ having mass $m$ and moment of inertia $I$ around the axis of precession. For the molecule $\alpha$, the linear speed ${\bf v}^\alpha(t)$ is related to the time rate of change of ${\bf r}^\alpha(t)$, and the corresponding 
angular speed $\bm{\omega}^\alpha(t)$ is related to the rate of change of ${\bf u}^\alpha$, through the relation (\ref{udot-e}). Here, we consider only this precession in the body fixed frame. The equations of motion for the linear and angular momenta variables $\{{\bf p}^\alpha,{\ell}^\alpha\}$ are obtained for the  Eqns. (\ref{m-lin})-(\ref{m-ang}).

The equations for the time evolution of these collective modes are obtained next. 
For thecollective variable of density defined in (\ref{def-rho}), the time derivatives are obtained in (\ref{def-g})-(\ref{def-l}) in terms of  the current densities, 
Thus the equation for the density $\hat{\rho}$ is obtained as,
\be \label{FP-r1} \frac{\pd\hat{\rho}({\bf x},t)}{\pd t} =
-\nabla_{\bf r}\sum_{\alpha=1}^N \dot{\bf x}^\alpha
\delta({\bf x}-{\bf x}^\alpha(t))-\nabla_{\bf u}\cdot \sum_{\alpha=1}^N \dot{\bf u}^\alpha\delta({\bf x}-{\bf x}^\alpha(t))~~. \ee
Using the definitions Eqn. (\ref{def-g})-(\ref{def-l}) we obtain:
\be \label{FP-r2} \frac{\pd\hat{\rho}({\bf x},t)}{\pd t} =
-\frac{1}{m}\nabla_{\bf r}\cdot\hat{\bf g}({\bf x},t) -\nabla_{\bf
    u} \cdot \sum_{\alpha} \{\bm{\omega}^\alpha(t)\times{\bf u}^\alpha(t)
\delta({\bf x}-{\bf x}^\alpha(t)) \}\ee
Using properties of the delta function and the definitions
(\ref{def-g})-(\ref{def-l}) for the currents, we obtain,
\be \label{FP-r31} \frac{\pd\hat{\rho}({\bf x},t)}{\pd t} =
-{m}^{-1}\nabla_{\bf r}\cdot\hat{\bf g}({\bf x},t)
-\rttensor{\bm{\kappa}}^{-1}\nabla_{\bf u} \cdot \{ \hat{\bm{\ell}}({\bf
    x},t){\times}{\bf u} \} \ee
In writing the second term on the right-hand side of Eq. (\ref{FP-r31}), we assume that for identical molecules, the principle moments of inertia in $\rttensor{\bm{\kappa}}_\alpha$ are independent of the label $\alpha$.
For the isotropic molecule using the set of principle axes for the moment of inertia $\rttensor{\bm{\kappa}}$, we have the diagonal form: ${\kappa}_{ij}=\delta_{ij}\kappa_0$ where $\kappa_0$ is the principle moment of inertia. The corresponding radius of gyration $l_g$ is obtained as $l_g^2=\kappa_0/m$. Eqn. (\ref{FP-r31}) is written as 
\be \boxed{\label{FP-r3} m\frac{\pd\hat{\rho}}{\pd t} 
    +\nabla_{\bf r}\cdot\hat{\bf g}
    +\frac{1}{l_g^2}\nabla_{\bf u} \cdot \{ \hat{\bm{\ell}}{\times}{\bf u} \}=0},\ee
where in order to avoid cluttering, we have dropped explicit mention of spatial and temporal dependence of the collective densities 
$\{\hat{\rho},\hat{\bf g},\hat{\bm{\ell}}\}$.
The equation of continuity with the rotational degree of freedom considered thus modifies to the form given by (\ref{FP-r3}).
From the definition (\ref{def-g}) the time derivative of momentum density $\hat{\bf
    g}({\bf x},t)$ is obtained as:
\bea \label{FP-g1} \frac{\pd{\hat{g}_i}}{\pd t} &=& \sum_{\alpha=1}^N
\Big [ -\nabla_{\bf r}^j \left \{ \frac{ p^\alpha_i
    {p^\alpha_j}}{m} \delta({\bf x}-{\bf x}^\alpha(t))\right
\}-\nabla_{\bf u}^j \left \{ p_i^\alpha\dot{\bf u}_j^\alpha
\delta({\bf x}-{\bf x}^\alpha(t))\right \}+ \dot{p}_i^\alpha
\delta({\bf x}-{\bf x}^\alpha(t))\Big ], ~~\nonumber\\
&=&   \sum_{\alpha=1}^N \Bigg [ -\nabla^j_{\bf r} \left \{
\frac{ p^\alpha_i{p^\alpha_j}}{m} \delta({\bf x}-{\bf x}^\alpha(t)) \right \}
-\nabla^j_{\bf u}  \left \{ p_i^\alpha\dot{\bf u}_j^\alpha
\delta({\bf x}-{\bf x}^\alpha(t))\right \}+ \nonumber \\
&+&\left (-\frac{p_i^\alpha}{m{\gamma_0}}+{\bm{f}}_i^\alpha \right ) \delta({\bf x}-{\bf x}^\alpha(t)) \Bigg ]+\hat{\eta}_i({\bf x},t) ~~.
\eea
In writing the second equality, we use the equation of motion (\ref{m-ang}).
The last term on the right-hand side of Eqn. (\ref{FP-g1}) represents the stochastic noise field defined in Eqn. (\ref{noise-r}).
Using the property of the delta function and  the equation of motion (\ref{udot-e}) for ${\bf u}^\alpha$ for the particle $\alpha$,
we obtain
\bea
\label{FP-g2}
\frac{\pd{\hat{g}_i}({\bf x},t)}{\pd t} &=& -\frac{1}{m{\gamma_0}}
{\hat{g}_i}({\bf x},t)+\hat{f}_i({\bf x},t)\hat{\rho}({\bf x},t)+ \hat{\eta}_i({\bf x},t) \\
&-& \sum_{\alpha=1}^N \Bigg [ \nabla_{\bf r}^j
\left \{\frac{ p^\alpha_i{p^\alpha_j}}{m} \delta({\bf x}-{\bf x}^\alpha(t)) \right \}
+\nabla^j_{\bf u}  \left \{ p_i^\alpha 
\epsilon_{jkm}{\omega}_k^\alpha{u}_m^\alpha
\delta({\bf x}-{\bf x}^\alpha(t))\right \} \Bigg ] ~~\nonumber
\eea
For an isotropic molecule with $\rttensor{\kappa}_{ij}=\delta_{ij}\kappa_0$, the last term is equal to, 
\bea \label{FP-g3}
\nabla^j_{\bf u}  \left \{ p_i^\alpha 
\epsilon_{jkm}{\omega}_k^\alpha{u}_m^\alpha
\delta({\bf x}-{\bf x}^\alpha(t))\right \}&=& {\kappa_0}^{-1}\sum_{\alpha=1}^N
{\bf u}\cdot\{\nabla_{\bf u} \delta({\bf x}-{\bf x}^\alpha(t)) \times{\ell}^\alpha\} p^\alpha_i \nonumber \\
&=&{\kappa_0}^{-1}\sum_{\alpha=1}^N
u_j\{\nabla_{\bf u}{\times}{\ell}^\alpha\delta({\bf x}-{\bf x}^\alpha(t))\}_j p^\alpha_i 
\eea
The equation for the momentum density $\hat{g}({\bf x},t)$ is thus obtained as
\be
\label{FP-g4}
\boxed{
    \frac{\pd{\hat{g}_i}}{\pd t} + \nabla_{\bf r}^j\hat{\sigma}_{ij}^{\rm gv}+({\bf u}\times \nabla_{\bf u})^j\hat{\sigma}_{ij}^{{\rm g}\omega}= -\frac{\hat{g}_i}{m{\gamma_0}}
    +\hat{f}_i\hat{\rho}+\hat{\eta}_i}
\ee
where  ${\sigma}^{\rm gv}_{ij}({\bf x},t)$ and ${\sigma}^{{\rm g}\omega}_{ij}({\bf x},t)$ in the right hand side of Eqn. (\ref{FP-g4}) are obtained as
\bea
\label{dCgv}
\hat{\sigma}_{ij}^{\rm gv}({\bf x},t) &=& m^{-1} \sum_{\alpha=1}^N 
\left \{p^\alpha_i{p^\alpha_j} \delta({\bf x}-{\bf x}^\alpha(t)) \right \} \\
\label{dCgo}
\hat{\sigma}_{ij}^{\rm g\omega}({\bf x},t) &=& {\kappa_0}^{-1}\sum_{\alpha=1}^N 
\left \{{{p^\alpha_i}{\ell}^\alpha_j}\delta({\bf x}-{\bf x}^\alpha(t)) \right \}~~.
\eea
Correlation of the noise $\hat{\bm{\eta}}(\bfx,t)$ is obtained in Eqn. (\ref{ncor-rr}).
Time derivative of $i$-th component of $\hat{\bm{\ell}}({\bf x},t)$ defined in Eqn. (\ref{def-l}) is obtained as,
\bea
\label{FP-el1}
&&\frac{\pd{\hat{\ell}_i}({\bf x},t)}{\pd t} = \sum_{\alpha=1}^N
\Bigg [ \dot{\ell}^\alpha_i -\nabla_{\bf r}^j{\ell}^\alpha_i \frac{p^\alpha_j}{m}  - \nabla_{\bf u}^j{\ell}^\alpha_i\dot{\bf u}^\alpha_j
\Bigg ] \delta({\bf x}-{\bf x}^\alpha(t)), \nonumber \\
\label{FP-el2}
&=&  \sum_\alpha \Bigg [\{-{\gamma}^{-1}_u
{\omega}_i^\alpha+{\tau}_i^\alpha+{\zeta}_i^\alpha\}
-\nabla_{\bf r}^j \frac{{\ell}^\alpha_j{p}_i^\alpha}{m}
-\nabla_{\bf u}^j{\ell}_i^\alpha{\{\bm{\omega}^\alpha\times{\bf u}^\alpha\}}_j \Bigg]
\delta({\bf x}-{\bf x}^\alpha).
\eea
Using the property of the delta function, we obtain
\be
\label{FP-el3}
\frac{\pd\hat{\ell}_i({\bf x},t)}{\pd t} = \hat{\cal C}_i({\bf x},t) -\frac{1}{{\gamma_u}\kappa_0} \hat{\ell}_i({\bf x},t) +\widehat{\tau}_i({\bf x},t)\hat{\rho}({\bf x},t) +\hat{\zeta}_i({\bf x},t) ~~. \ee
The first term on the right-hand side of Eqn. (\ref{FP-el3})  is obtained  as
\bea
\label{def-bell}
{\cal C}_i({\bf x},t)&=&-m^{-1}\nabla_{\bf r}^j \Bigg [ \sum_{\alpha=1}^N
{p_i^\alpha{\ell}^\alpha_j}\delta({\bf x}-{\bf x}^\alpha)\Bigg]
+ {\kappa_0}^{-1}({\nabla_{\bf u} \times {\bf u}})_j \Bigg [ \sum_{\alpha=1}^N
{\ell_i^\alpha}{\ell_j^\alpha}\delta({\bf x}-{\bf x}^\alpha)\Bigg]~~. \nonumber \\
&=&-\nabla_{\bf r}^j \hat{\sigma}_{ij}^{\ell{\rm v}}({\bf x},t) 
+mT ({\nabla_{\bf u} \times {\bf u}})_j \hat{\sigma}_{ij}^{\ell\omega}({\bf x},t). ~~
\eea
where we have adopted a notation similar to (\ref{dCgv})-(\ref{dCgo}) to define,
\bea
\label{dClv}
\hat{\sigma}_{ij}^{\ell{\rm v}}({\bf x},t) &=& m^{-1}\sum_{\alpha=1}^N 
\left \{ {p^\alpha_i}{\ell}^\alpha_j\delta({\bf x}-{\bf x}^\alpha(t)) \right \}~~,\\
\label{dClo}
\hat{\sigma}_{ij}^{\ell\omega}({\bf x},t) &=& {\kappa_0}^{-1}\sum_{\alpha=1}^N 
\left \{ {\ell^\alpha_i}{\ell}^\alpha_j\delta({\bf x}-{\bf x}^\alpha(t)) \right \}~~.
\eea
The function $\widehat{\bm{\tau}}({\bf x},t)$ in the third term on the
right-hand side of Eqn. (\ref{FP-el3}) 
is defined in terms of the torque $\bm{\tau}^\alpha$ on the particle $\alpha$, 
in Eqns. (\ref{def-tor}). 
Finally, the last term on the right-hand side of Eqn. (\ref{FP-el3}) 
is the stochastic field defined in Eqn. (\ref{noise-z}).
The correlation of this noise field $\hat{\bm{\zeta}}(\bfx,t)$ is obtained using
the microscopic relations  (\ref{tor-cor}) as,
\be 
\label{ncor-zeta} \langle \hat{\zeta}_i({\bf x} ,t)
\hat{\zeta}_j({\bf x}^\prime,t^\prime) \rangle{=}
2mT{\gamma_u^{-1}}\delta_{ij}\hat{\rho}({\bfx},t)\delta({\bf x}-{\bf x}')\delta(t-t')~~.
\ee
Hence the balance equation for angular momentum density $\hat{\bm{\ell}}({\bf x},t)$ is obtained as,
\be
\label{FP-el3f}
\boxed{\frac{\pd\hat{\ell}_i}{\pd t} +\nabla_{\bf r}^j \hat{\sigma}_{ij}^{\ell{\rm v}} +({\nabla_{\bf u} \times {\bf u}})^j \hat{\sigma}_{ij}^{\ell\omega}
    =  -{({\gamma_u}I)}^{-1} \hat{\ell}_i +\widehat{\tau}_i\hat{\rho}
    +\hat{\zeta}_i ~~.  } \ee
The Fokke-Planck dynamics is described with the stochastic differential equations (\ref{FP-r3}), (\ref{FP-g4}), and (\ref{FP-el3f}) for the time evolutions of the respective collective densities $\{\hat{\rho}(\bfx,t),\hat{\bf g}(\bfx,t),\hat{\bm{\ell}(\bfx,t)}\}$.
These three equations represent the rotational and translational Brownian dynamics of the fluid molecules described by Eqs. (\ref{m-lin})-(\ref{def-tor}).

\section{Coarse graining: Field theoretic model}

The equations of hydrodynamics involve the time evolution of a set of local densities, which generally \cite{nambu} reflects the underlying conservation laws of the system. The microscopic densities introduced in earlier sections are defined as a sum over the Dirac delta functions such that integration over a small volume around a chosen point obtains the content of a corresponding physical entity (mass, momentum, energy) in that volume. The definitions of mass, linear momentum,  angular momentum, and energy density are respectively given in Eqns. (\ref{def-rho})-(\ref{cons-edef}). 
The corresponding hydrodynamic equations for the {\em smoothly varying} local densities are respectively denoted by $\{ {\rho} ({\bf x},t), {\bf g} ({\bf x},t),  {\ell}({\bf x},t),{e}({\bf x},t)\}$ are obtained by taking average of the
above balance equations. The averaging, in principle, needs to be
done over the nonequilibrium ensemble. If the system is in
equilibrium, averaging the densities obtains the corresponding
time-independent equilibrium value. The nonequilibrium average is
obtained by extending the notion of the Gibbsian ensembles \cite{mcquarie} to
include systems out of equilibrium. More specifically, the
nonequilibrium system that we describe here is in
the time regime where it is assumed to have reached a state of {\em local
    equilibrium}, {\em i.e.}, the set of local
densities $\{\hat{a} ({\bf x})\}$ are sufficient to describe the
state of the system. This is a plausible hypothesis, particularly
at high densities, since the stage of local equilibrium is reached
rapidly through frequent inter-particle collisions.

\subsection{Local Equilibrium distribution}

The distribution function represents the probability of finding the given system in certain parts of the phase space. We denote the phase point $\Gamma_N$ by the full set  phase space variables, consisting of the configuration-variables ${\bf x}^\alpha\equiv\{{\bf r}^\alpha,{\bf u}^\alpha\}$  as well as the momentum variables
$\{{\bf p}^\alpha,\bm{\ell}^\alpha\}$ ($\alpha=1,..N$).
The probability function for the local-equilibrium state is obtained
in analogy with the {\em true-equilibrium}
state in terms of a set of extensive conserved quantities $ {\bf A} = \{H,N, {\bf P}_0,{\bf L}_0...\}$, which respectively denote the total energy, number of particles, the total linear momentum and the total angular momentum about the centre of mass (say) for the system. These extensive properties are dependent on the phase space variables $\Gamma_N$. For the true equilibrium  state, the canonical distribution function is
\be 
\label{loceqdis1}
f_{eq}(\Gamma_N) \  \sim \exp [ - \beta (H-\mu N + {\bf v}_0.{\bf
    P}_0+\omega_0.{\bf L}_0)]~~.
\ee
The corresponding thermodynamic variables constituting the set $\{\beta,\mu,{\bf v}_0,\bm{\omega}_0\}$ are intensive and respectively define
the temperature, chemical potential, linear-velocity, and angular-velocity fields in the fluid. Analogously, the {\em local} equilibrium state of the fluid is
described by the nonuniform densities $\hat{\bf a}({\bf r})
\equiv \{\hat{e},\hat{\rho},\hat{\bf g},\hat{\ell}\}$. The functions $\{\hat{a}({\bf r})\}$. The local densities corresponding to a set of conserved properties $\{{\cal A}\}$ 
are,
\be
\label{Le-conden} {\cal A}= \int d{\bf r} \hat{a}({\bf r})~~.
\ee
$\{\alpha_a\}$ is the corresponding local thermodynamic property
(intensive) in terms of which the local equilibrium is defined.
The distribution function is obtained as,
\be
\label{le-d} f_{le} ({\Gamma}_N,t) = Q_l^{-1} {\exp}\Big \{ -
\int d{\bf x} \beta\Big [ \{\hat{e}({\bfx})-{\bf v}.\hat{\bf
    g}({\bf x})-{\bf \omega}.\hat{\ell}({\bf x})
-\tilde{\mu}\hat{\rho}({\bf x}) \Big ] \Big \}, 
\ee
In analogy with the equilibrium case, the set of local
thermodynamic variables are given by, $\{\beta({\bf x},t),{\mu}({\bf x},t),{\bf v}({\bf x},t),\bm{\omega}({\bf x},t)\}$ respectively
referring to the local temperature, local chemical potential,
local velocity, and local angular velocity. 
$Q^{-1}_l$ is the necessary normalization constant which therefore
satisfies the relation
\be
\label{Le-part} Q_l = \mathrm{Tr} \left [ \exp\Big \{ - \int d{\bf
    x} \sum_{\{a\}} \alpha_a ({\bf x},t) \hat{a}({\bf x}) \Big \}
\right ]~~,
\ee
where ``Tr" denotes the integration over all the phase space variables.
The quantity $\tilde{\mu}({\bfx},t)$ in
the right hand side of (\ref{le-d}) is defined as,
\be \label{tilde-mu1} \tilde{\mu}({\bf x},t)=\mu ({\bf x},t)-\frac{1}{2} m v^2 ({\bf x},t) - \frac{1}{2}\{\bm{\omega}({\bf x},t) \cdot \rttensor{\bm{\kappa}}\cdot\bm{\omega}({\bf x},t)\} \ee
where $\rttensor{\bm{\kappa}}$ is  the moment of inertia tensor for the $\alpha$-molecule (assumed independent of $\alpha$).

The nonuniform parameters characterizing the
local equilibrium (LE) are determined by imposing the self
consistency condition that the nonequilibrium state average of a
local density $\hat{a}$ can be approximated by performing the
average over the local equilibrium distribution.
\be
\label{self} a({\bf x},t) = {<\hat{a}({\bf x})>}_{n.e} =
{<\hat{a}({\bf x})>}_{l.e}.
\ee
The local ensemble average of $\hat{a}$ is obtained from the
relation
\be
\label{Le-avden} {<\hat{a}({\bf x})>}_{le} = -\frac{\delta\ln
    Q_l(t)}{\delta\alpha_a({\bf x},t)}
\ee
In the form considered in eqn. (\ref{le-d}) we, therefore, have the
following $\alpha_a$'s corresponding to the set $\hat{a} \equiv
\{\hat{e},\hat{\bf g},\hat{\ell},\hat{\rho}\}$
\bea
\label{Le-loctherm} \alpha_n &=& -\beta({\bf x},t) \tilde{\mu}
({\bf x},t), \ \ \ \
\alpha_e = \beta({\bf x},t)\\
\alpha_{\bf g} &=& -\beta({\bf x},t) {\bf v}({\bf x},t) , \ \ \ \
\alpha_{\ell}= -\beta({\bf x},t)\omega({\bf x},t)~~.\nonumber
\eea

\subsection{Comoving frame}

We consider a transformation of the microscopic variables for the system to a comoving frame with local velocity  ${\bf v}({\bfx},t)$ and angular velocity $\bm{\omega}(\bfx,t)$. In addition to the local frame having a translational velocity ${\bf v}({\bf x},t)$, it also has a local angular speed $\omega({\bf x},t)$ at the point ${\bf x}{\equiv}\{{\bf r},{\bf u}\}$ at time $t$.
\bea \label{ct1} {\bf x}^\alpha &=& {\bf x}^{\prime\alpha}, \ \ \  {i.e.},\  \ 
\{{\bf r}^\alpha = {\bf r}^{\prime\alpha}, 
{\bf u}^\alpha = {\bf u}^{\prime\alpha}\}, ~~\\
\label{ct2}
{\bf p}^\alpha &=& {\bf p}^{\prime\alpha}+m{\bf v}({\bf x}^\alpha,t),~~~\\
\label{ct3}
\bm{\omega}^\alpha &=& \bm{\omega}^{\prime\alpha}+
\bm{\omega}({\bf x}_\alpha,t)~,
\eea
It is useful to note here that the  transformation (\ref{ct3})
follows simply if the comoving frame with angular
speed $\omega({\bf x}^\alpha,t)$ in the same direction as
$\bm{\omega}^\alpha$ for the $\alpha$-th molecule.
The relation between the angular momenta $\bm{\ell}^\alpha$ in the lab and comoving frame is obtained using the definition for $\bm{\ell}^\alpha$ in terms of the intra-molecular coordinates,
\bea \bm{\ell}^\alpha &=& \sum_\nu {\bf r}_\nu^{\prime\alpha} \times
m^\alpha_\nu {\bf v}_\nu^{\prime\alpha} \\
&=& \sum_{\nu\prime} 
{\bf r}_\nu^{\prime\alpha} \times m^\alpha_\nu \bm{\omega}^\alpha
\times {\bf r}_\nu^{\prime\alpha}=\rttensor{\bm{\kappa}}^\alpha \bm{\omega}^\alpha 
\eea
The relation between $\bm{\ell}^\alpha$ in the primed and unprimed frames is therefore obtained as,
\be
\label{angtran}
\bm{\ell}^\alpha = \bm{\ell}^{\prime\alpha}+\rttensor{\bm{\kappa}}\bm{\omega}({\bf x}_\alpha)~~.
\ee
For the energy density, by taking a dot product of the vector ${\bf p}^\alpha$ with itself, we obtain using (\ref{ct2}) the relation,
\be \label{tran-eT} \frac{{\bf p}^\alpha \cdot {\bf
        p}^\alpha}{2m}  = \frac{{\bf p}^{\prime\alpha} \cdot {\bf
        p}^{\prime\alpha}}{2m} + {\bf p}^{\prime\alpha} \cdot {\bf v}({\bf
    x}^\alpha) - \frac{m}{2}{v}^2({\bf x}^\alpha)~~.\ee
Multiplying Eqn. (\ref{tran-eT}) with $\delta({\bf x}-{\bf x}^\alpha)$ 
and summing over all $\alpha$ we obtain for the
translational part $e^\mathrm{T}$, defined in Eqn. (\ref{e-tran}), the following 
transformation rule:
\be \label{tran-eT1} \hat{e}_\mathrm{T}({\bf x},t) =
\hat{e}_\mathrm{T}^\prime({\bf x},t) + \hat{\bf g}({\bf x},t)\cdot
{\bf v}({\bf x},t) - \frac{m}{2}{v}^2({\bf x},t) \hat{\rho}({\bf
    x},t) ~~.\ee
Next, with a dot product of the vector $\bm{\ell}^\alpha$ 
with $\rttensor{\bm{\kappa}}^{-1}$ from both right and left side, we obtain,
\be \label{ad1}
\bm{\ell}^\alpha \cdot \rttensor{\bm{\kappa}}^{-1} \cdot \bm{\ell}_\alpha = \bm{\ell}^\prime_\alpha \cdot \rttensor{\kappa}^{-1}\cdot \bm{\ell}^\prime_\alpha+ \bm{\omega}({\bf x}_\alpha)\cdot\rttensor{\kappa}\cdot\bm{\omega}({\bf x}_\alpha) + 
2\bm{\ell}^{\prime\alpha}\cdot\bm{\omega}({\bf x}^\alpha)~~. \ee
Multiplying Eqn. (\ref{ct2}) and (\ref{ct3}) with $\delta({\bf x}-{\bf
    x}^\alpha)$ and summing over all $\alpha$ we obtain respectively,
\bea \label{tran-vR1} \hat{\bf g}({\bf x},t) &=&
\hat{\bf g}^\prime({\bf x},t) + m\hat{\rho}({\bf x},t){\bf v}({\bf x},t) , \\
\label{tran-vR2} \hat{\bm{\ell}}({\bf x},t) &=&
\hat{\bm{\ell}}^\prime({\bf x},t) + \hat{\rho}({\bf x},t)
\rttensor{\bm{\kappa}}\bm{\omega}({\bf x},t).~~
\eea
Next, multiplying Eqn. (\ref{ad1}) with $\delta({\bf x}-{\bf
    x}^\alpha)$ and summing over all $\alpha$ we obtain,
for the rotational energy $e^\mathrm{R}$, defined in Eqn. (\ref{e-rot}),
\be \label{tran-eR1} \hat{e}_\mathrm{R}({\bf x},t) =
\hat{e}_\mathrm{R}^\prime({\bf x},t) + \bm{\ell}^\prime({\bf x},t)\cdot
\bm{\omega}({\bf x},t)+\frac{1}{2}\Big \{\bm{\omega}({\bf
    x},t)\cdot\rttensor{\bm{\kappa}}\cdot\bm{\omega}({\bf x},t) \Big \}\hat{\rho}({\bf
    x},t) ~~.\ee
Eq. (\ref{tran-eR1}) defines the transformation for energy density.

We now average over the local equilibrium distribution the linear equations
(\ref{tran-vR1}), and (\ref{tran-vR2}). For
the primed quantities of the local rest frame vanished since the distribution has a ${\bf p}^{\prime\alpha} \leftrightarrows- {\bf p}_\alpha^\prime$ and also ${\ell}_\alpha^\prime \leftrightarrows - {\ell}_\alpha^\prime$ symmetry. We obtain,
\bea \label{tran-R1} {\bf g}({\bf x},t) &=& m{\rho}({\bf x},t){\bf v}({\bf x},t) , \\
\label{tran-R2}\bm{\ell}({\bf x},t) &=& \rho({\bf x},t)
\rttensor{\bm{\kappa}}\bm{\omega}({\bf x},t).~~
\eea
Using eqns. (\ref{le-d}), (\ref{tilde-mu1}), (\ref{tran-eT1}), and
(\ref{tran-eR1}), we obtain that the local equilibrium
distribution in terms of the primed quantities reduces to the Gibsian
form for the Grand Canonical ensemble:
\be \label{le-dprime} f_{le} ({\Gamma}_N,t) = Q_l^{-1} {\exp}\Big
\{ - \int d{\bf r} \beta({\bf x},t) \Big \{\hat{e}^\prime ({\bf
    x}) -{\mu}({\bf x},t)\hat{\rho}({\bf x})\Big \}. \ee
In the primed coordinates, interpreting total energy density $ \hat{e}^\prime({\bf x})=\hat{e}^\prime_\mathrm{T}({\bf x})+
\hat{e}^\prime_\mathrm{R}({\bf x})$, the fluid is locally at rest, and the corresponding probability distribution reduces to the grand-canonical form. Note that $\mu({\bf x})$ depends on both position ${\bf r}$ and angular coordinates 
${\bf u}$.
.

\section{Coarse-grained equations}

\subsection{Smoluchowski dynamics; In terms of \{x\}}

To obtain an equation for the coarse-grained density $\rho(\bfx,t)$ when the particles are following the overdamped or the Smoluchowsky type dynamics,  the microscopic equation for the collective density for
$\hat{\rho}(\bfx,t)$ is averaged over the local equilibrium distribution. First, we consider the microscopic equations Eqn. (\ref{BD-comp}) and (\ref{j-total}) corresponding to Statonvitch interpretation of the noise in Eq. (\ref{e0-u}), and treat the corresponding case of I\^{t}o noise at the end of this section. The local equilibrium averages below are $\langle ...\rangle$. Reorganizing the terms on the right-hand side of Eq. (\ref{j-total}), the averaged equation is obtained as:
\bea
\label{BD-cre4}&&
\frac{\pd}{\pd t}{\rho}({\bf x},t)
= {\gamma_0}\nabla_{\bf r}^i
\Big [ T\nabla_{\bf r}^i{\rho}({\bf x},t) -\Big \{ \Big \langle\hat{\bm{f}}_i({\bf x},t)
\hat{\rho}({\bf x},t)\Big \rangle+{\eta}_i({\bf x},t) \Big \} \Big ]  \\
&+& {\gamma_u}\Big [
T\{\nabla^i_{\bf u}{\Pi}_{ij}(u) \nabla^j_{\bf u}\}
{\rho}({\bf x},t) - \nabla_{\bf u}^i \Big \{
\Big \langle \widehat{\bm{\tau}}({\bf x},t)\times{\bf u}\Big \}_i \hat{\rho}({\bf x},t)
\Big \rangle
+\nabla_{\bf u}^i \{ {\zeta}({\bf x},t) \times {\bf u} \}_i \Big ]~. \nonumber
\eea
The noises $\eta({\bf x},t)$ and $\zeta({\bf x},t)$ are the respective local equilibrium averages of the two noises $\hat{\eta}({\bf x},t)$ and $\hat{\zeta}({\bf x},t)$ appearing in Eqs. (\ref{noise-r}) and \ref{noise-z}). The two averages involving the force density
$\hat{\bm{f}}({\bf x},t)$ and the torque-density $\widehat{\bm{\tau}}({\bf x},t)$ is determined in the local equilibrium distribution in the primed coordinates 
introduced in Eq. (\ref{ct1})-(\ref{ct3}). To express $\{\hat{\bm{f}}({\bf x},t),\widehat{\bm{\tau}}({\bf x},t)\}$ in terms of the local densities we introduce an operator ${\cal L}$ which is defined in terms of the phase space coordinates
as follows:
\bea \label{liou-op}
i{\cal L}[{\bf x}, {\bf p},\bm{\ell}]&=&\sum_\alpha \left \{ 
\frac{\pd{H}}{\pd{\bf p}^\alpha}\cdot\frac{\pd}{\pd{\bf r}^\alpha}-
\frac{\pd{H}}{\pd{\bf r}^\alpha}\cdot\frac{\pd}{\pd{\bf p}^\alpha} +
\frac{\pd{H}}{\pd{\bm{\ell}}^\alpha}\cdot  \left (  
{\bf u}^\alpha \times \frac{\pd}{\pd{\bf u}^\alpha} \right ) 
- \left (
{\bf u}^\alpha \times \frac{\pd{H}}{\pd{\bf u}^\alpha} \right ) \cdot\frac{\pd}{\pd{\bm{\ell}}^\alpha}
\right \}  \nonumber \\
&=&
\label{liou-op1}
\sum_\alpha \left \{ 
{\bf v}^\alpha\cdot\nabla_{{\bf r}^\alpha}+ {\bm{f}}^\alpha\cdot\nabla_{{\bf p}^\alpha} + 
\dot{\bf u}^\alpha\cdot\nabla_{{\bf u}^\alpha}+ \bm{\tau}^\alpha\cdot\nabla_{\bm{\ell}^\alpha} 
\right \}
\eea
The force ${\bm{f}}^\alpha$ and the torque $\bm{\tau}^\alpha$ respectively appearing in the second and fourth term on the right-hand side of Eq. (\ref{liou-op1}) are as defined in Eqs. (\ref{def-for}) and (\ref{def-tor}).
We use the relation (\ref{udot-e}) in writing the third term on the 
right-hand side of Eq. (\ref{liou-op1}). Unlike the Liouvillian operator, acting ${\cal L}$ on the momentum density 
${\bf g}(\bfx,t)$ does not give rise to the corresponding equation of motion since stochastic and dissipative parts are also involved. $i{\cal L}H=0$ by construction. We work with the operator ${\cal L}$ in terms of the primed coordinates defined in Eq. (\ref{ct1})-(\ref{ct3}). Applying $i{\cal L}[{\bf x}, {\bf p}^\prime,\bm{\ell}^\prime]$ on $\rho(\bfx,t)$ we obtain, using (\ref{udot-e}) 
\bea
\label{iLrho}
i{\cal L}[{\bf x}, {\bf p}^\prime,\bm{\ell}^\prime]\hat{\rho}({\bf x}) 
&=&  -\sum_\alpha \Big [
\nabla_{\bf r}\cdot \{ {\bf p}^{\prime\alpha}
\delta({\bf x}-{\bf x}^\alpha(t))\} +\nabla_{\bf u} \cdot \{ 
\dot{\bf u}^\alpha \delta({\bf x}-{\bf x}^\alpha(t)) \} \Big ]~~, \nonumber \\
&=&  - 
\nabla_{\bf r}\cdot\hat{\bf g}^\prime(\bfx,t) -
m\nabla_{\bf u} \cdot \{ \rttensor{\bm{\kappa}}^{-1} \hat{\bm{\ell}}^\prime(\bfx,t)
\times {\bf u} \} \nonumber \\
&=& -\nabla_{\bf r}\cdot\hat{\bf g}^\prime(\bfx,t) -
l_0^{-2}\nabla_{\bf u} \cdot \{ \hat{\bm{\ell}}^\prime(\bfx,t)
\times {\bf u} \}
\eea
Similarly, applying  $i{\cal L}[{\bf x}, {\bf p}^\prime,\bm{\ell}^\prime]$ on $\hat{\bf g}^\prime({\bf x},t)$, we obtain,
\bea
\label{iLg}
&&i{\cal L}\hat{g}^\prime_i({\bf x}) = i{\cal L}[{\bf x}, {\bf p}^\prime,\bm{\ell}^\prime]\sum_\alpha {\bf p}^{\prime\alpha}_i \delta({\bf x}-{\bf x}^\alpha(t))\\
&=& \sum_\alpha \Big [ {\bm{f}}_i^\alpha \delta({\bf x}-{\bf x}^\alpha(t)) 
-\nabla_{\bf r}\cdot \{ m{\bf v}^{\prime\alpha}{\bf v}^{\prime\alpha}_i
\delta({\bf x}-{\bf x}^\alpha(t))\} -\nabla_{\bf u} \cdot \{ {\bf p}^{\prime\alpha}_i
\dot{\bf u}^{\alpha} \delta({\bf x}-{\bf x}^\alpha(t)) \} \Big ]~~. \nonumber 
\eea
The microscopic variables $\dot{\bf r}^{\prime\alpha}={\bf p}^{\prime\alpha}/m={\bf v}^{\prime\alpha}$ and $\bm{\omega}^{\prime\alpha}$ are  independent. In terms of the primed variables, the local equilibrium distribution is symmetric under ${\bf p}^{\prime\alpha} \leftrightarrows -{\bf p}^{\prime\alpha}$ and $\bm{\omega}^{\prime\alpha} \leftrightarrows -\bm{\omega}^{\prime\alpha}$. The corresponding local equilibrium average of  $i{\cal L}{\bf g}^\prime_i({\bf x})$ with distribution (\ref{le-dprime})  obtains
\be
\label{lg1}
\Big \langle i{\cal L}\hat{g}^\prime_i({\bf x})\Big \rangle
= \Big \langle \hat{\bm{f}}_i\hat{\rho}({\bf x})\Big \rangle -T\nabla_{\bf r}^i \rho({\bf x})~~.
\ee
From now on, we will consider the system to be at a fixed temperature so that $\beta(\bfx,t)=\beta$, and the primary focus is on the role of the local chemical potential. The left-hand side of  (\ref{lg1}) is evaluated, making use of the derivative form of $i{\cal L}$, and acting it on the distribution function $f_{\rm l}(\Gamma^\prime)$.
Integrating by parts and using $i{\cal L}\mathrm{H}=0$,
we obtain:
\bea  &&{\langle -i{\cal L} \hat{g}^\prime_i({\bf x})
    \rangle}= -\int d\Gamma \hat{g}^\prime_i({\bf x})
\frac{1}{Q} i{\cal L} {\exp}\Big [ -\beta \{ \mathrm{H}-\int d{\bf
    x}^\prime \mu ({\bf x}^\prime)\hat{\rho}({\bf x}^\prime) \} \Big ]
\nonumber \\ 
&=& \beta\int d{\bf x}^\prime  \mu ({\bf x}^\prime)
{\langle \hat{g}^\prime_i({\bf x})i{\cal L}\hat{\rho} ({\bf
        x}^\prime)\rangle }\nonumber \\
\label{liou-g} &=& -\beta\int d{\bf x}^\prime  \mu ({\bf
    x}^\prime)\Big [ \sum_j {\langle \hat{g}^\prime_i({\bf x})
    {\nabla}_j^\prime \hat{g}^\prime_j ({\bf x}^\prime) \rangle
}-\frac{1}{l_0^2}\sum_{j,k,m}
\nabla^j_{\bf u} \epsilon_{jkm}
{\langle \hat{g}^\prime_i({\bf x})\hat{\ell}_k^\prime(\bfx^\prime)\rangle}
{u}_m \} \Big ]~~.\eea
The average in the second term on the right-hand side of Eq. (\ref{liou-g}) vanished since in the primed variables the local equilibrium distribution is symmetric under ${\bf p}^{\prime\alpha} \leftrightarrows -{\bf p}^{\prime\alpha}$ and $\bm{\omega}^{\prime\alpha} \leftrightarrows -\bm{\omega}^{\prime\alpha}$.  This substitution obtains the simple relation:
\be
\label{lg2}
\Big \langle i{\cal L}\hat{\bf g}^\prime_i({\bf x})\Big \rangle
=  \Big \langle \hat{\rho}({\bf x})\Big \rangle \nabla^i_{\bf r} \mu({\bf x}))=
{\rho}({\bf x})\nabla^i_{\bf r} \mu({\bf x}))~~~.
\ee
Using the results (\ref{lg2}) and (\ref{lg1}) we obtain,
\be
\label{lg3}
\Big \langle \hat{\bm{f}}_i\hat{\rho}({\bf x})\Big \rangle
= T\nabla_{\bf r}^i \rho({\bf x})+ {\rho}({\bf x})\nabla^i_{\bf r} \mu({\bf x}))~~.
\ee
In a similar way, the fifth term on the RHS of Eqn. (\ref{BD-cre4}) involving the torque-density $\widehat{\bm{\tau}}({\bf x},t)$ is calculated. Acting $i{\cal L}[{\bf x}, {\bf p}^\prime,\bm{\ell}^\prime]$ on $\hat{\ell}^\prime({\bf x},t)$, we obtain
\bea
\label{iL1}
&&i{\cal L}\hat{\ell}^\prime_i({\bf x}) = i{\cal L}\sum_\alpha {\ell}^{\prime\alpha}_i \delta({\bf x}-{\bf x}^\alpha(t))\\
&=& \sum_\alpha \Big [ {\tau}^\alpha_i \delta({\bf x}-{\bf x}^\alpha(t)) 
-\nabla_{\bf r}\cdot \{{\ell}^{\prime\alpha}_i{\bf v}^{\prime\alpha}
\delta({\bf x}-{\bf x}^\alpha(t))\} -\nabla_{\bf u} \cdot \{ {\ell}^{\prime\alpha}_i
{\dot{\bf u}}^\alpha \delta({\bf x}-{\bf x}^\alpha(t)) \} \Big ]~~. \nonumber 
\eea
The microscopic variables  ${\bf p}^{\prime\alpha}$ and $\bm{\omega}^\alpha$ are treated as  independent. Using the symmetries of the Hamiltonian under the transformations, ${\bf p}^{\prime\alpha} \leftrightarrows -{\bf p}^{\prime\alpha}$ and $\bm{\omega}^{\prime\alpha} \leftrightarrows -\bm{\omega}^{\prime\alpha}$
in the primed (comoving) frame, the corresponding
local equilibrium average  $\braket{{\cal L}{\ell}^\prime_i({\bf x})}$ with distribution (\ref{le-d}), is obtained as
\be
\label{iL2}
\braket{i{\cal L}\hat{\ell}^\prime_i({\bf x})} =\braket{\widehat{\tau}_i ({\bf x},t)\hat{\rho}({\bf x},t)} 
-\nabla_{\bf u} \cdot \Big \langle {\ell}^{\prime\alpha}_i
{\dot{\bf u}}^{\prime\alpha}\delta({\bf x}-{\bf x}^\alpha(t))  \Big \rangle ~~. 
\ee
The last term on the right-hand side, with substitution of $\dot{\bf u}^{\prime\alpha}=\bm{\omega}^{\prime\alpha}\times{\bf u}^{\prime\alpha}$ reduces to,
\bea
\label{iL3}
\nabla_{\bf u}^j \Big \langle {\ell}^{\prime\alpha}_i
{(\bm{\omega}^{\prime\alpha}\times{\bf u}^{\prime\alpha})}_j \delta({\bf x}-{\bf x}^\alpha(t)) \Big \rangle
&=& \nabla_{\bf u}^j \Big \langle \rttensor{\bm{\kappa}}_{ik}\omega^{\prime\alpha}_k\epsilon_{jmn}\omega^{\prime\alpha}_m u^\alpha_n \delta({\bf x}-{\bf x}^\alpha(t)) \Big \rangle.
\eea
The average $\langle \rttensor{\bm{\kappa}}_{jk}\omega^{\prime\alpha}_k\omega^{\prime\alpha}_m \rangle$
is with respect to the Gibbsian distribution (\ref{le-d})  with Hamiltonian
$H^\prime$ which involves  $\frac{1}{2}\omega^{\prime\alpha}_i\rttensor{\bm{\kappa}}_{ij}\omega^{\prime\alpha}_j$ in the rotational
part of the kinetic energy term. It follows then,
$\rttensor{\bm{\kappa}}_{jk}\langle \omega^{\prime\alpha}_k\omega^{\prime\alpha}_m \rangle=T\delta_{jm}$.
Substituting this in Eqn. (\ref{iL3}), we write the result (\ref{iL2}) as,
\be
\label{iL4}
\braket{i{\cal L}\hat{\bm{\ell}}^\prime({\bf x})} =\braket{\widehat{\bm{\tau}} ({\bf x},t)\hat{\rho}({\bf x},t)} -T\{  {\bf u} \times\nabla_{\bf u}\} {\rho}({\bf x},t)~~,
\ee
since $\nabla_{\bf u} \times {\bf u}=0$.  Vector multiplication of Eq. (\ref{iL4}) with ${\bf u}$ obtains for the $i$-th component,
\bea
\label{iL5}
\braket{i{\cal L}\hat{\bm{\ell}}^\prime({\bf x})}\times{\bf u}  &=& 
\braket{\widehat{\bm{\tau}}({\bf x},t)\hat{\rho}({\bf x},t)}\times{\bf u}+T{\bf u} 
\times \{ {\bf u} \times\nabla_{\bf u} \rho({\bf x},t)\},\nonumber \\
&=& \braket{\widehat{\bm{\tau}}({\bf x},t)\hat{\rho}({\bf x},t)}\times{\bf u} 
-T\{\rttensor{\bm{\Pi}}\nabla_{\bf u} \}\rho({\bf x},t),~~.
\eea
where the tensor $\rttensor{\bm{\Pi}}$ is as defined in Eq. (\ref{Pi-def}) . Using these results for the force-density and torque-density, the 
Eqn. (\ref{BD-cre4}) for coarse-grained density reduces to the form:
\bea
\label{BD-cre5}&&
\frac{\pd}{\pd t}{\rho}({\bf x},t)
= {\gamma_0}\bm{\nabla}_{\bf r}\cdot
\Big [ \rho({\bf x},t) \bm{\nabla}_{\bf r}\mu({\bf x})
-\bm{\eta}({\bf x},t) \Big \} \Big ]  \\
&+& {\gamma_u}\Big [
T\{\nabla^i_{\bf u}{\Pi}_{ij}(u) \nabla^j_{\bf u}\}
{\rho}({\bf x},t) -\bm{\nabla}_{\bf u}\cdot\Big \{
\Big \langle \widehat{\bm{\tau}}({\bf x},t)\hat{\rho}({\bf x},t)\Big \rangle\times{\bf u}\Big \}
-\bm{\nabla}_{\bf u} \cdot \{ \bm{\zeta}({\bf x},t) \times {\bf u} \}  \Big ] \nonumber
\eea
Eqn. (\ref{BD-cre5}) for time evolution of the coarse-grained $\rho({\bf x},t)$ is further simplified to obtain:
\be
\label{BD-cr6}
\frac{\pd}{\pd t}{\rho}({\bf x},t) = {\gamma_0}\bm{\nabla}_{\bf r} \cdot \Big [ \rho({\bf x},t) \bm{\nabla}_{\bf r} \mu({\bf x}) \Big ]  + {\gamma_u} \bm{\nabla}_{\bf u} \cdot\Big [\braket{i{\cal L}\hat{\ell}^\prime({\bf x})}\times{\bf u}\Big ] +{\theta}({\bf x},t)~.
\ee
The noise term ${\theta}({\bf x},t)$ in Eqn. (\ref{BD-cr6}) is obtained as
\be 
\label{BDc3}
{\theta}({\bf x},t)=-{\gamma_0}\bm{\nabla}_{\bf r}\cdot
\bm{\eta}({\bf x},t) - {\gamma_u}\bm{\nabla}_{\bf u}\cdot
\{\bm{\zeta}({\bf x},t)\times{\bf u}\}~~. \ee
Eqn. (\ref{BDc3}) is a local equilibrium Eqn average. (\ref{tot-noise}) for the microscopic equations.
Using the derivative form of ${\cal L}$,  and the local equilibrium distribution Eq.~(\ref{le-d})  for the phase space variables $\Gamma_N^\prime$,  we obtain the second term on the right-hand side of Eqn. (\ref{BD-cr6}) as,
\bea
\label{aman1}
&&\braket{i{\cal L}\hat{\ell}_i^\prime({\bf x})} 
= -\int d\Gamma^\prime_N \hat{\ell}_i^\prime({\bf x}) Q^{-1}
i{\cal L} e^{-\beta H^\prime+\int d{\bf x}
    \mu({\bf x})\hat{\rho}({\bf x})} \nonumber \\
&=& -\int d\Gamma^\prime_N\int d{\bf x}_1 \mu({\bf x}_1)\{i{\cal L} \hat{\rho}({\bf x}_1)\} \hat{\ell}_i^\prime({\bf x})
Q^{-1}e^{-\beta H^\prime+\int d{\bf x}\mu({\bf x})\hat{\rho}({\bf x})} \nonumber \\
&=& -\int d{\bf x}_1 \mu({\bf x}_1)\braket{\hat{\ell}_i^\prime({\bf x})
    \{i{\cal L} \hat{\rho}({\bf x}_1)\}}~~ . 
\eea
We have used the result $i{\cal L}H^\prime=0$, in getting the second equality in (\ref{aman1}).
The definitions (\ref{def-rho}) and (\ref{liou-op}) and, respectively, for the operator $\hat{\rho}$, and ${\cal L}$ are used to obtain the local equilibrium average appearing in the right-hand side of (\ref{aman1}) as,
\bea
\label{aman2}
&&\braket{i{\cal L}\hat{\ell}_i^\prime({\bf x})}
= -\int d{\bf x}_1 \mu({\bf x}_1)\braket{\hat{\ell}_i^\prime({\bf x})
    \Big \{ -m^{-1}\nabla_{{\bf r}_1}\hat{\bf g}^\prime({\bf x}_1)
    -\nabla_{{\bf u}_1}\cdot(\rttensor{\bm{\kappa}}^{-1}\bm{\hat{\ell}}^\prime({\bf x}_1)\times{\bf u}_1) \Big \}},
\nonumber \\
&=& \int d{\bf x}_1 \mu({\bf x}_1)
\Big \{ m^{-1}\nabla_{{\bf r}_1}\cdot\braket{ \hat{\ell}_i^\prime({\bf x})\hat{\bf g}^\prime({\bf x}_1)}
+\nabla_{{\bf u}_1}\cdot\rttensor{\bm{\kappa}}^{-1}\braket{\hat{\ell}_i^\prime({\bf x})\bm{\hat{\ell}}^\prime({\bf x}_1)}\times{\bf u}_1) \Big \}.
\eea
Assuming that $\bm{\hat{\ell}}^\prime({\bf x})$ and ${\bf g}^\prime({\bf x})$ are independent, the average $\braket{ \hat{\ell}_i^\prime({\bf x})\hat{\bf g}^\prime({\bf x}_1)}$ goes to zero  since in the primed (comoving) frame the corresponding
local equilibrium distribution have ${\bf p}^{\prime\alpha} \leftrightarrows -{\bf p}^{\prime\alpha}$ and ${\omega}^{\prime\alpha} \leftrightarrows -{\omega}^{\prime\alpha}$
symmetry. For the second term on the right-hand side of Eqn. (\ref{aman2}) reduces to the form,
\bea
\label{aman3}
\Big \{ \nabla_{{\bf u}_1}\cdot\rttensor{\bm{\kappa}}^{-1}\braket{\hat{\ell}^\prime_i({\bf x})\bm{\hat{\ell}}^\prime({\bf x}_1)}\times{\bf u}_1) \Big \} =  \nabla^p_{{\bf u}_1} \epsilon_{pjl}
\rttensor{\bm{\kappa}}_{jk}^{-1}\braket{\hat{\ell}_i^\prime({\bf x})\hat{\ell}_k^\prime({\bf x}_1)}{u}_{1l}) \Big \} 
\nonumber
\eea
Using the following result for the average over local equilibrium distribution, 
\be
\label{ll-av}
\rttensor{\bm{\kappa}}_{jk}^{-1}\braket{\hat{\ell}_i^\prime({\bf x})\hat{\ell}_k^\prime({\bf x}_1)}=\delta_{ji}
\delta({\bf x}-{\bf x}_1)\braket{\hat{\rho}({\bf x})}~.\ee
Since $\nabla_{\bf u}\times{\bf u}=0$ we obtain the result
\bea
\label{aman4}
&&\braket{i{\cal L}\hat{\ell}_i^\prime({\bf x})}
= \int d{\bf x}_1 \mu({\bf x}_1)
\nabla^p_{{\bf u}_1} \epsilon_{jlp}\delta({\bf x}-{\bf x}_1)\delta_{ji}
{u}_{1l} \Big \} \rho({\bf x}) \nonumber \\
&=&\int d{\bf x}_1\mu({\bf x}_1)  \{{\bf u}_1 
\times \nabla_{{\bf u}_1} \}_i \left [ \rho({\bf x})\delta({\bf x}-{\bf x}_1)\right ] 
\nonumber \\
&=&-\int d{\bf x}_1\left [ \rho({\bf x})\delta({\bf x}-{\bf x}_1)\right ] 
\{{\bf u}_1  \times \nabla_{{\bf u}_1} \}_i \mu({\bf x}_1) =
-\rho({\bf x})\{{\bf u} \times \nabla_{\bf u} \}_i \mu({\bf x})
\eea
Vector multiplication of ${\bf u}$ yields,
$\{ \braket{i{\cal L}\hat{\ell}_i^\prime({\bf x})}\times {\bf u} \}_i
= \rho({\bf x})\left \{\Pi_{ij}(u) \nabla^j_{\bf u}\right \} \mu({\bf x})$. Using this
result, Eq. (\ref{BD-cr6}) reduces to: 
\be
\label{BD-cr7}
\frac{\pd}{\pd t}{\rho}({\bf x},t) = {\gamma_0}\delta_{ij} \{\nabla^i_{\bf r} \rho({\bf x}) \nabla^j_{\bf r} \}\mu({\bf x}) + {\gamma_u} \{ \bm{\nabla}_{\bf u}^i 
\Pi_{ij}(u) \rho({\bf x}) \bm{\nabla}^j_{\bf u} \} \mu({\bf x})  
+{\theta}({\bf x},t) ~~.
\ee
The above equation for the coarse-grained density $\rho({\bf x},t)$ is written in a more compact form of a stochastic partial differential equation (PDE):
\be
\label{DK-stat}
\boxed{\frac{\pd}{\pd t}{\rho}({\bf x},t)
    = \Big \{ \nabla_{\bf x}^i {\cal D}_{ij}
    [{\rho}({\bf x},t)]\nabla_{\bf x}^j  \Big \}{\frac{\delta{F}[{\rho}]}{\delta{\rho}(\bfx,t)}}
    {+} {\theta}({\bf x},t)}~,
\ee
with $\mu({\bf x})=\left ({\delta{F}[{\rho}]}/{\delta{\rho}}\right )$  the chemical potential introduced in defining the local equilibrium distribution. 
The field $\theta(\bfx,t)$ in the last term on the right-hand side of Eqn. (\ref{DK-stat}) represent a stochastic part of the dynamics and is obtained as a local equilibrium average of the noise $\hat{\theta}(\bfx,t)$.
The correlation of the noise $\theta$ averaged over the same local equilibrium ensemble\cite{pre2013}, is obtained using Eqns.
(\ref{ncor-rr})-(\ref{ncor-uu})  as,
\be
\label{BDc4}
\langle{\theta}({\bf x} ,t){\theta}({\bf x}^\prime,t) \rangle =
2T\Big \{ \nabla_{\bf x}^i {\cal D}_{ij} [{\rho}]\nabla_{\bf x}^j  \Big \}
\delta({\bf x}-{\bf x}')\delta(t-t') ~~.
\ee
The noise $\theta$ is therefore multiplicative due to the presence of $\rho(\bfx,t)$ on the right-hand side of (\ref{BDc4}).
The transport matrix ${\cal D}$ has a $2d\times{2d}$ block diagonal form as 
defined in Eq. (\ref{diffu-matrix}). 

Adopting the I\^{t}o interpretation of the noise in the microscopic Eqn. (\ref{e0-u}) for the microscopic equations Eqn. (\ref{BD-ito}) and (\ref{j-total}) must be averaged over the local equilibrium distribution. An extra term ${-2\bm{\nabla}_{\bf u}}{\cdot}({\bf u}\hat{\rho}(\bfx,t))$ occurs in the right hand side of (\ref{BD-ito}), as compared to the corresponding equation (\ref{BD-stat}) for the previous case of Stratonovich interpretation. Since this contribution is linear in the density field $\hat{\rho}$, the local equilibrium average of this term is straightforward forward, and we obtain the corresponding equation for the coarse-grained density field $\rho(\bfx,t)$ as,
\be
\label{DK-ito}\boxed{
    \frac{\pd}{\pd t}{\rho}({\bf x},t)
    {=}\left \{\nabla_{\bf x}^i \widetilde{\cal D}_{ij}{\rho}(\bfx,t)\nabla_{\bf x}^j\right \}
    {\frac{\delta{F}[{\rho}]}{\delta{\rho}(\bfx,t)}}{+}
    \left \{\nabla_{\bf x}^i {\cal D}^0_{ij}{\rho}(\bfx,t)\nabla_{\bf x}^j\right \}
    {\frac{\delta{F}^\prime[{\rho}]}{\delta{\rho}(\bfx,t)}}   
    +{\theta}({\bf x},t) ~~.}
\ee
The functional ${F}^\prime[\rho]$ introduced on the right hand side of Eq. (\ref{BD-re5-ito})  is obtained as,
\be \label{fen-ito}{F}^\prime[\rho]{=} {F}[\rho]{-}{d}{\int}{d}{\bf x} {{\rho}({\bf x})}
{\ln {u}}.
\ee
From Eq. (\ref{fen-ito}), it follows that for the dynamics which maintain the
magnitude of $u$ fixed, the difference between  
${F}$ and ${F}^\prime$ is a constant, assuming the total number of particles remains unchanged. It is useful to note here that the basic equation of motion for the microscopic variable ${\bf u}^\alpha$ ( See Eqns. (\ref{ma-ang}) and (\ref{e0-u})) requires that its magnitude remains unchanged in time.


\subsection{Fokker Planck dynamics: In terms of \{x,p\}}

In this case, the local equilibrium distribution, as discussed with Eqn. (\ref{le-d}) in section II is obtained as:
\be
f_{\rm le}(\Gamma_N;t)
= Q^{-1}\exp \Big (-\beta H+\int d{\bf x}
\beta \left [ \tilde{\mu}({\bf x})\hat{\rho}({\bf x})
+\hat{\bf g}({\bf x})\cdot{\bf v}({\bf x}) +\hat{\ell}({\bf x})
{\omega}({\bf x}) \right] \Big ) ,
\ee
where ${\ell}({\bf x},t)$ is the local angular momentum field required to include rotational freedom. We average Eqs. (\ref{FP-r3}),(\ref{FP-g4}), and (\ref{FP-el3f}) respectively to obtain the following coarse-grained equations,
\bea
\label{fpc-den} &&m\frac{\pd{\rho}({\bf x},t)}{\pd t} 
{+}\nabla_{\bf r}{\cdot}{\bf g}({\bf x},t)
{+}\frac{1}{l_g^2}\nabla_{\bf u}{\cdot}\{{\ell}(\bfx,t){\times}{\bf u} \}{=}0~, \\
\label{fpc-mom}
&&\frac{\pd{g_i}({\bf x},t)}{\pd t}{+}\nabla_{\bf r}^j\braket{\hat{\sigma}_{ij}^{\rm gv}}
{+}({\bf u}\times \nabla_{\bf u})^j\braket{\hat{\sigma}_{ij}^{{\rm g}\omega}}{=}
\frac{{g_i}({\bf x},t)}{m{\gamma_0}}
{+}\braket{\hat{\bm{f}}_i({\bf x},t)\hat{\rho}({\bf x},t)}
{+}\eta_i({\bf x},t), \\
\label{fpc-ell}
&&\frac{\pd{\ell}_i({\bf x},t)}{\pd t}{+}\nabla_{\bf r}^j \braket{\hat{\sigma}_{ij}^{\ell{\rm v}}}{+}({\nabla_{\bf u} \times {\bf u}})^j \braket{\hat{\sigma}_{ij}^{\ell\omega}}
{=}\frac{\ell_i(\bfx,t)}{\gamma_u{\kappa_0}}
{+}\braket{\widehat{\tau}_i({\bf x},t)\hat{\rho}({\bf x},t)}
{+}\zeta_i({\bf x},t) ~~.  \eea
The noises $\bm{\eta}({\bf x},t)$ and $\bm{\zeta}({\bf x},t)$ in right hand sides of Eqn. (\ref{fpc-mom}) and Eqn. (\ref{fpc-ell}) are respectively the local equilibrium averages of the two noises $\hat{\eta}({\bf x},t)$ and $\hat{\zeta}({\bf x},t)$ defined in Eqs. (\ref{noise-r}) and (\ref{noise-z}). Correlations of the noises $\eta(\bfx,t)$ and $\zeta(\bfx,t)$ are obtained as
\bea \label{cg-rr} \langle {\eta}_i({\bf x} ,t)
{\eta}_j({\bf x}^\prime,t^\prime) \rangle &=&
2mT{\gamma_0^{-1}}\delta_{ij}{\rho}({\bf
    x},t)\delta({\bf x}-{\bf x}')\delta(t-t')\\
\label{cg-ll}
\langle{\zeta}_i({\bf x} ,t)
{\zeta}_j({\bf x}^\prime,t^\prime) \rangle&=&
2mT{\gamma_u^{-1}}\delta_{ij}{\rho}({\bfx},t)\delta({\bf x}-{\bf x}')\delta(t-t')~~.
\eea
We simplify the Eqn. (\ref{fpc-mom}) by evaluating  $\braket{\hat{\sigma}_{ij}^{\rm gv}}$ and $\braket{\hat{\sigma}_{ij}^{{\rm g}\omega}}$ using their respective definitions (\ref{dCgv}) and (\ref{dCgo}).
Next, the transformations (\ref{ct1})-(\ref{ct3}) for the momentum ${\bf p}^\alpha$ and angular speed ${\omega}^\alpha$ are applied in (\ref{dCgo}). We use the fact that in the local comoving frame the averages of $\{{\bf p}^{\prime\alpha},{\omega}^{\prime\alpha}\}$, both vanish due to the  ${\bf p}^{\prime\alpha} \leftrightarrows -{\bf p}^{\prime\alpha}$ and ${\omega}^{\prime\alpha} \leftrightarrows -{\omega}^{\prime\alpha}$ symmetries of the Hamiltonian $H^\prime$.  Hence we obtain
\bea
\label{dCgv1}
\braket{\hat{\sigma}_{ij}^{\rm gv}({\bf x},t)} &=&  {\bf v}_i({\bf x},t){\bf v}_j({\bf x},t)m\rho({\bf x},t)
+T\delta_{ij}m\rho({\bf x},t)~, \\
\label{dCgo1}
\braket{\hat{\sigma}_{ij}^{\rm g\omega}({\bf x},t)}&=& {\bf v}_i({\bf x},t){\omega}_j({\bf x},t)m\rho({\bf x},t)~~.
\eea
The thermodynamic fields ${\bf v}({\bf x},t)$ and $\omega({\bf x},t)$ are replaced in terms of the coarse grained densities $\{\rho({\bf x},t),{\bf g}({\bf x},t),{\ell}({\bf x},t)\}$ using Eqs. (\ref{tran-R1}) and (\ref{tran-R2}).  Here, we have used the inertia tensor $\rttensor{\bm{\kappa}}$ to be diagonal. Finally, using results (\ref{dCgv1}), (\ref{dCgo1}) and (\ref{lg1}) in Eqn. (\ref{fpc-mom}) we obtain the equation for coarse-grained momentum density in the form:
\bea
\label{mom-e1}
&&\frac{\pd{g_i}({\bf x},t)}{\pd t}{+}\frac{1}{m}\nabla_{\bf r}^j \left [ \frac{g_i({\bf x},t)g_j({\bf x},t)}{\rho({\bf x},t)} \right ]{+}
\frac{1}{\kappa_0}\widetilde{\nabla}_{\bf u}^j 
\left [\frac{g_i({\bf x},t)\ell_j({\bf x},t)}{\rho({\bf x},t)}\right ]
{+}\frac{{g_i}({\bf x},t)}{{\gamma_0}m}\nonumber \\
&=&\braket{i{\cal L}\hat{g}^\prime_i({\bf x})}
+\eta_i({\bf x},t),
\eea
where we have introduced, in the abbreviated form, the operator $\widetilde{\nabla}_{\bf u}{=}{\bf u}{\times}\nabla_{\bf u}$. 
Next, we simplify the Eqn. (\ref{fpc-ell}) evaluating  $\braket{\hat{\sigma}_{ij}^{\ell{\rm v}}}$ and $\braket{\hat{\sigma}_{ij}^{\ell{\omega}}}$ using their respective definitions
(\ref{dClv}) and (\ref{dClo}). We apply in Eqn. (\ref{dClo}) the transformations (\ref{ct1})-(\ref{ct3}) 
for the momentum ${\bf p}^\alpha$ and angular speed ${\omega}^\alpha$. In the  local co-moving frame the averages of $\{{\bf p}^{\prime\alpha},{\omega}^{\prime\alpha}\}$, both vanish due to the  ${\bf p}^{\prime\alpha} \leftrightarrows -{\bf p}^{\prime\alpha}$ and ${\omega}^{\prime\alpha} \leftrightarrows -{\omega}^{\prime\alpha}$ symmetries of the Hamiltonian $H^\prime$.  We obtain
\bea
\label{dClv1}
\braket{\hat{\sigma}_{ij}^{\ell{\rm v}}({\bf x},t)} &=& 
{\kappa_0}{\bf v}_i({\bf x},t){\omega}_j({\bf x},t)\rho({\bf x},t), \\
\label{dClo1}
\braket{\hat{\sigma}_{ij}^{\ell{\omega}}({\bf x},t)} &=& {\ell}_i({\bf x},t){\omega}_j({\bf x},t)~
+T\delta_{ij}I\rho({\bf x},t)~~.
\eea
The thermodynamic fields ${\bf v}({\bf x},t)$ and $\omega({\bf x},t)$ are replaced in terms of the coarse grained densities $\{\rho({\bf x},t),{\bf g}({\bf x},t),{\ell}({\bf x},t)\}$ witht the use of Eqns. (\ref{tran-R1}) and (\ref{tran-R2}). Finally, using results (\ref{dClv1}), (\ref{dClo1}) and (\ref{iL4}) in Eqn. (\ref{fpc-ell}) we obtain the equation for coarse grained momentum density $\bm{\ell}(\bfx,t)$ in the form:
\bea
\label{ell-e1}
\frac{\pd{\ell}_i({\bf x},t)}{\pd t}&+&\frac{1}{m}\nabla_{\bf r}^j \left [ \frac{g_i({\bf x},t)\ell_j({\bf x},t)}{\rho({\bf x},t)} \right ]+\frac{1}{\kappa_0}({\bf u}{\times}\nabla_{\bf u})^j \Big [ \frac{\ell_i({\bf x},t)\ell_j({\bf x},t)}{\rho({\bf x},t)}\Big ] =  \\ 
&-& {(\gamma_u{\kappa_0})}^{-1}{\ell_i({\bf x},t)}+
\braket{\widehat{\tau}_i({\bf x},t)\hat{\rho}({\bf x},t)}-T({\bf u}\times \nabla_{\bf u})^i
\rho({\bf x},t)+\zeta_i({\bf x},t) ~~.  \nonumber \\
&=&-{(\gamma_u{\kappa_0})}^{-1}{\ell_i({\bf x},t)}+\braket{i{\cal L}\hat{\ell}^\prime_i({\bf x})}+\zeta_i({\bf x},t) ~~.  \nonumber
\eea
Eqns. (\ref{mom-e1}) and (\ref{ell-e1}) for the coarse-grained variables ${\bf g}({\bf x},t)$ and  $\ell({\bf x},t)$ are very similar and they require evaluation of the respective averages, $\braket{i{\cal L}\hat{g}^\prime_i({\bf x})}$, and $\braket{i{\cal L}\hat{\ell}^\prime_i({\bf x})}$ to reach a closed form.
These two local equilibrium averages are evaluated in Eqns, respectively. (\ref{lg2}) and (\ref{aman4}), obtaining the following FNH equations for the set of variables 
$\{\rho,{\bf g},\bm{\ell}\}$:
\bea
\label{den-fnl}&&
\frac{\pd{\rho}}{\pd t}
{+}\frac{1}{m}\nabla^j_{\bf r}g_j{+}\frac{1}{\kappa_0}\widetilde{\nabla}^j_{\bf u}{\ell}_j{=}0~, \\
\label{mom-fnl}&&
\frac{\pd{g_i}}{\pd t}{+}\frac{1}{m}\nabla_{\bf r}^j \left [ \frac{g_ig_j}{\rho} \right ]{+}
\frac{1}{\kappa_0}\widetilde{\nabla}_{\bf u}^j 
\left [\frac{g_i\ell_j}{\rho}\right ]{+}\rho \nabla^i_{\bf r}{\mu}
{+}\frac{g_i}{{\gamma_0}m}{=}\eta_i~,  \\
\label{ell-fnl}&&
\frac{\pd{\ell}_i}{\pd t}{+}\frac{1}{m}\nabla_{\bf r}^j \left [ \frac{g_i\ell_j}{\rho} \right ] {+}\frac{1}{\kappa_0}\widetilde{\nabla}_{\bf u}^j \left [ \frac{\ell_i\ell_j}{\rho}\right ]
{+}\rho\widetilde{\nabla}_{\bf u}^i \mu{+}
\frac{\ell_i}{\gamma_u{\kappa_0}}{=}\zeta_i~~.
\eea
Eqn. (\ref{den-fnl}) is the continuity equation representing reversible dynamics.
Eqns. (\ref{mom-fnl}) and (\ref{ell-fnl}) takes a more familiar form using 
the classical density functional\cite{tvr} expression for the chemical potential
$\mu({\bf x})$ in terms of the functional derivative of the free energy treated as a functional $F_U$ of density only:
 \be
  \label{cpot-der} 
 \mu({\bfx}){=}\frac{{\delta}{F_U[\rho]}}{\delta\rho(\bfx)}~.
         \ee
 Under time reversal, for both Eqn. (\ref{mom-fnl}) and (\ref{ell-fnl}) the respective left-hand sides, the first four terms remain invariant, while the fifth term changes sign, signifying irreversibility in dynamics. The dissipative terms in both equations are linear in the fields. The noise $\bm{\eta}$ and $\bm{\zeta}$ respectively appearing in the two equations are both multiplicative, as can be seen from the respective noise correlations (\ref{cg-rr}) and (\ref{cg-ll}).

\section{Equations of  FNH: Poisson Brackets}
    
The equations of fluctuating nonlinear hydrodynamics depicted above for the set of collective variables $\psi_a{\in}\{\rho,{\bf g},\bm{\ell}\}$ are expressed in the 
well known Time Dependent Ginzburg Landu (TDGL) \cite{Halperin-rmp} form:
\be 
\label{tdgl-eq}
\frac{\pd\psi_a}{\pd{t}}+V^a[\psi]+L^0_{ab}\frac{\delta{\cal F[\psi]}}{\delta{\psi_b}}=\theta_a,
\ee
where $V^a$ denotes the reversible part of the equation of motion and $L_{ab}^0$ signifies the $ab$ element of the bare transport matrix.  $\theta_a$ is the stochastic part or the noise and is multiplicative in the present case. The reversible part of the dynamics, {\em i.e.}, $V^a$ for the  collective variable $\psi_a$
is obtained in the form 
\be \label{rev-part} V^a= \sum_b Q_{ab}[\psi]\frac{\delta{\cal F[\psi]}}{\delta{\psi_a}}~.\ee
The functional ${\cal F}[\psi]$ in the right-hand side of Eqn. (\ref{rev-part}) is expressed in terms of the coarse-grained variables $\psi$. The stationary solution of this Fokker- Planck equation corresponding to the generalized Langevin equation (\ref{tdgl-eq}) is obtained as $\exp[-{\cal F}]$, and hence ${\cal F}$ is termed as the free energy functional in the field-theoretic model. 
The elements of matrix $Q_{ab}$ are obtained from the classical Poisson Brackets (PB) \cite{volvick}, $\hat{Q}_{ab}(\bfx,\bfx^\prime)\equiv\{\hat{\psi}_a(\bfx),\hat{\psi} _b(\bfx^\prime)\}$ by replacing the microscopic collective variables (hatted) in the latter with their respective coarse-grained (unhatted) counterparts. The PB between dynamical (microscopic) variables $A$ and $B$ are defined as follows:
\be
\label{pb-def0}
\{A,B\}= \sum_{\alpha=1}^N \Bigg [ \Big (
\bm{\nabla}_{{\bf r}^\alpha}A\cdot\bm{\nabla}_{{\bf p}^\alpha}B
-\bm{\nabla}_{{\bf r}^\alpha}B\cdot\bm{\nabla}_{{\bf p}^\alpha}A
\Big ){+}
\Big (
\widetilde{\bm{\nabla}}_{{\bf u}^\alpha}A\cdot\bm{\nabla}_{{\bm{\ell}}^\alpha}B
- \widetilde{\bm{\nabla}}_{{\bf u}^\alpha}B\cdot\bm{\nabla}_{{\bm{\ell}}^\alpha}A
\Big ) \Bigg ]~~.
\ee
By definition, the PBs satisfy the antisymmetry relation, $Q_{ab}=-Q_{ba}$. Following the above definition, the PBs between the phase space variables in the present case are obtained as follows:
\bea
\label{pblu}
\{u_i^\alpha,\ell_j^\nu\}&=&\delta_{\alpha\nu}\epsilon_{ijp}u^\alpha_p \\
\label{pbrp}
\{r_i^\alpha,p_j^\nu\}&=&\delta_{\alpha\nu}
\eea
Using the definition (\ref{pb-def0}) the PB's between the collective variables $\hat{\psi}_a{\equiv}\{
\hat{\rho},\hat{\bf g},\hat{\bm{\ell}}\}$ are obtained in Appendix \ref{app2} in Eqs. (\ref{rho-rho})-(\ref{ell-ell}). Using the results for the PB, the reversible terms $V^a[\psi]$ ( see eqn. (\ref{rev-part}) ) for the respective fields are obtained as follows. 
\bea
\label{v_rho} 
V_\rho&=& -\frac{1}{m}\nabla_{\bf r}^i{g_i}
-\frac{1}{\kappa_0}\widetilde{\nabla}_{\bf u}^i {\ell_i} \\
\label{v_g} 
V^i_{g}&=& -\rho\nabla_{\bf r}\frac{\delta{\cal F}_U}{\delta{\rho}}{-}
\nabla_{\bf r}^j\left [ \frac{g_i{g_j}}{\rho} \right ]{-}\nabla_{\bf r}^j\left [
\frac{g_i{\ell_j}}{\rho} \right ] ~~,\\
\label{v_ell} 
V^i_{\ell}&=& 
-{\rho}\widetilde{\nabla}_{\bf u}^i
\frac{\delta{\cal F}_U}{\delta{\rho}}-\nabla_{\bf r}^j\Big [ 
\frac{{\ell_i}{g_j}}{\rho} \Big ]
-\widetilde{\nabla}_{\bf u}^j \Big [ \frac{{\ell_i}{\ell_j}}{\rho} \Big ]~~.
\eea
Details of these calculations are presented in Appendix \ref{app2}. The irreversible part of the equation of motion is $L^0_{ab}({\delta{\cal F}}/{\delta\psi_b})$ whcih involves the bare transport matrix  $L_{ab}^0$. The bare transport matrix is symmetric by choice $L_{ab}^0$=$L_{ba}^0$. The elments like $L^0_{\rho\psi}=0$ for  $\psi\in\{\rho,g_i,\ell_i\}$,  and continuity equation is maintained (\ref{den-fnl}). The matrix elements $L_{g_i,g_j}^0$, $L_{\ell_i,\ell_j}^0$ are chosen in diagonal form, and $L^0_{g_i\ell_j}=0$. With this choice for the bare-dissipation matrix, we obtain the equations (\ref{den-fnl})-(\ref{ell-fnl}) of FNH in the TDGL form of Eq. (\ref{tdgl-eq}). The corresponding Fokker-Planck equation describes the time evolution of the probability distribution ${\cal P}[\rho,{\bf g},\bm{\ell};t]$, and the stationary solution is obtained as $\exp \{ -{\cal F}[\rho,{\bf g},\bm{\ell}] \}$ \cite{ma-mazenko}. The model involves only the density variable $\rho$; the latter follows Langevin dynamics (\ref{DK-stat}) or (\ref{DK-ito}) with only the dissipative term. The reversible part vanishes in this case since $Q_{\rho\rho}$ vanishes by definition. The corresponding Fokker Planck equation describes the time evolution of the probability distribution ${\cal P}[\rho;t]$, and the stationary solution for ${\cal P}$ is $\exp \{ -{\cal F}[\rho] \}$, with ${\cal F}[\rho]$ signifying the free energy functional expressed in terms of the coarse-grained density  $\rho$ only.

\subsection{Free energy functional of coarse-grained variables}

Next, we calculate the free energy functional for the system in terms of the coarse-grained fields $\{\hat{\bf a}\}$. First, the free energy is obtained from the corresponding local equilibrium distribution $f_{le}(\Gamma_N; \alpha_a(\bfx,t))$,
\be
\label{f0-fp}
\widetilde{\cal F}[\alpha_a]
= -k_B{T}\ln\int d\Gamma_N f_{le}({\Gamma}_N; \alpha_a(\bfx,t)),
\ee
which is a functional of the conjugate thermodynamic fields $\{\alpha_a\}$.  
\be
\label{Le-part1} f_l ({\Gamma}_N; \alpha_a(\bfx,t))= \frac{1}{Q_l(\alpha_a)} \exp\Big \{ - \int d{\bf
    x} \sum_{\{a\}} \alpha_a ({\bf x},t) \hat{a}({\bf x}) \Big \}~~,
\ee
where $Q_l$ is the necessary normalization constant, such that
\be
\label{Le-part2} Q_l = \mathrm{Tr} \left [ \exp\Big \{ - \int d{\bf
    x} \sum_{\{a\}} \alpha_a ({\bf x},t) \hat{a}({\bf x}) \Big \}
\right ]~~.
\ee
The correpsonding free energy is obtained as $\widetilde{\cal F}[\alpha_a]=-k_BT\ln Q_l$.

In the Fokker-Planck description for the dynamics of the fluid molecules with translational and rotational motions, the set of microscopic densities is $\hat{\bf a}({\bf r})\equiv \{\hat{e},\hat{\rho},\hat{\bf g},\hat{\ell}\}$, and the corresponding local thermodynamic fields $\{\alpha_i\}\equiv\{\beta(\bfx,t),\tilde{\mu}(\bfx,t),{\bf v}(\bfx,t),\bm{\omega}(\bfx,t)\}$. The  distribution function $ f_{le}({\Gamma}_N,t)$ is obtained in Eq. (\ref{le-d}).  The corresponding free energy functional 
$\widetilde{\cal F}[\tilde{\mu}(\bfx),{\bf v}(\bfx),\bm{\omega}(\bfx)]$ is defined as:
\be
\label{f2-fp}
\widetilde{\cal F}[\tilde{\mu},{\bf v},\bm{\omega}]
= -k_B{T}\ln\int d\Gamma_N f_{le}({\Gamma}_N,t; \tilde{\mu},{\bf v},\bm{\omega}) 
\ee
The free energy ${\cal F}[\rho,{\bf g},\bm{\ell}]$ 
in terms of the set of coarse-grained fields  $\{{\rho}(\bfx),{\bf g}(\bfx),\bm{\ell}(\bfx)\}$,is obtained through a Legendre transoformation:
\be
\label{f1-fp} 
{\cal F}[\rho,{\bf g},\bm{\ell}]
=\widetilde{\cal F}[\tilde{\mu},{\bf v},\bm{\omega}]
+\int d\bfx \Big[ {\rho}(\bfx)\tilde{\mu}(\bfx)
+{\bf g}(\bfx).{\bf v}(\bfx)+\bm{\ell}(\bfx).\bm{\omega}(\bfx) \Big ] 
\ee
Next, we consider the expression for the free energy of the fluid with Smoluchowsky dynamics. The {\em local} equilibrium state is
characterized by the nonuniform densities $\hat{\bf a}({\bf r})
\equiv \{\hat{e},\hat{\rho}\}$ with the distribution function,
\be
\label{le-dd1} f^0_{le} ({\Gamma}_N,t;\mu(\bfx,t)) = Q_l^{-1} {\exp}\Big \{ -
\int d{\bf x} \beta({\bf x},t) \Big [ \{\hat{e}_\mathrm{T}({\bf
    x})-{\mu}({\bf x},t)\hat{\rho}({\bf x}) \Big ] \Big \}.
\ee
The free energy functional in this case $\widetilde{\cal F}_U[\mu({\bf x})]$ is obtained as
\be
\label{sm-pf-fe}
\widetilde{\cal F}_U[\mu(\bfx)]
= -\kB T\ln\int d\Gamma_N f^0_{le}({\Gamma}_N,t;\mu(\bfx,t)) 
\ee
The phase space variable $\Gamma_N$ in this case is only over the configuration variables ${\bf x}^\alpha\equiv\{{\bf r}^\alpha.{\bf u}^\alpha\}$ ( $\alpha=1,....N$).
The corresponding free energy ${\cal F}_U[{\rho}(\bfx)]$ in terms of the coarse grained fileds  $\rho(\bfx)$ is obtained through a Legendre transoformation as:
\be
\label{sm-fen}
{\cal F}_U[{\rho}(\bfx)]
= \widetilde{\cal F}_U[\mu(\bfx)]+\int d\bfx{\rho}(\bfx)\mu(\bfx),
\ee
In order to obtain $\widetilde{\cal F}[\tilde{\mu}(\bf x),{\bf v}(\bf x),\bm{\omega}({\bf x})]$ defined 
in Eq. (\ref{f2-fp}), we calculate the corresponding integration with respect to the momentum and angular momentum and obtain for fixed (inverse) temperature $\beta$,
\bea
\label{fp-fe1}
\widetilde{\cal F}[\tilde{\mu}(\bf x),{\bf v}(\bf x),\bm{\omega}({\bf x})] &=& -\kB T
\ln\int d{\bf x}^N\exp \Big [ -\beta V+\beta\int d{\bfx}\rho(\bfx)\tilde{\mu}(\bfx) 
\nonumber \\
&+& \frac{\beta}{2}\int d{\bfx} \rho(\bfx)\Big \{ m{\bf v}^2(\bfx)
+\kappa_0\bm{\omega}^2(\bfx) \Big \} \Big ]
\eea
where $V$ is the potential energy part of the Hamiltonian, introduced in Eq. (\ref{pote}). Here, we have ignored an ideal gas contribution, which is a constant for fixed temperature. On the other hand, from Eq. (\ref{le-dd1}), and (\ref{sm-fen}), we obtain the free energy  corresponding to the Smoluchosky dynamics as,
\be
\label{sm-pf-fe1}
\widetilde{\cal F}_U[\mu(\bfx)]
= -\kB T\ln{\int}d{\bf x}^N\exp \Big [ -\beta V+\beta\int d{\bfx}\rho(\bfx){\mu}(\bfx) ~.
\ee
Now, comparing Eqs. (\ref{fp-fe1}) and (\ref{sm-pf-fe1}), it follows
\be
\label{link-fe}
\widetilde{\cal F}[\tilde{\mu}(\bfx),{\bf v}(\bfx),\bm{\omega}(\bfx)]
=\widetilde{\cal F}_U \left[\tilde{\mu}(\bfx)+\frac{m}{2}{\bf v}^2(\bfx)+\frac{\kappa_0}{2}
{\bm{\omega}^2(\bfx)} \right] .
\ee
Substitution of Eq. (\ref{link-fe}) into right hand side if Eq. (\ref{f1-fp}) yields
\bea
\label{f1-fp1} 
&&{\cal F}[\rho,{\bf g},\bm{\ell}]
={\cal F}_U \left[ {\rho}\right]{-}\int{d\bfx}{\rho}(\bfx)\mu(\bfx)
{+}\int{d\bfx} \Big[ {\rho}(\bfx)\tilde{\mu}(\bfx)
{+}{\bf g}(\bfx).{\bf v}(\bfx){+}\bm{\ell}(\bfx).\bm{\omega}(\bfx) \Big ]  \nonumber \\
&=&{\cal F}_U \left[ {\rho}\right]{+}
\int{d{\bfx}} \left [- {\rho}(\bfx) \left \{ \frac{m}{2}{\bf v}^2(\bfx)+\frac{\kappa_0}{2}
\bm{\omega}^2(\bfx) \right \}
+{\bf g}(\bfx).{\bf v}(\bfx)+\bm{\ell}(\bfx).\bm{\omega}(\bfx) \right ] . \nonumber \\
\eea
Now using the relations (\ref{tran-R1})-(\ref{tran-R2}) between the coarse-grained momenta and the corresponding thermodynamic fields, the second term on the right-hand side of Eq (\ref{f1-fp1}) simplifies to obtain the following relation between the
free energy functional in the Fokker-Planck and Smoluchosky descriptions.
\bea
\label{two-reln}
{\cal F}[\rho,{\bf g},\bm{\ell}]
&=& {\cal F}_U \left[\rho\right] 
+\int d\bfx \Bigg [ \frac{{\bf g}^2(\bfx)}{2m\rho(\bfx)}+
\frac{\bm{\ell}^2(\bfx)}{2\kappa_0\rho(\bfx)} \Bigg ]{\equiv}
{\cal F}_U \left[\rho\right] +{\cal F}_K \left[\rho,{\bf g},\bm{\ell}\right] \\
\label{one-rein}
{\cal F}[\rho,{\bf g}]
&=& {\cal F}_U \left[\rho\right] 
+\int d\bfx \Bigg [ \frac{{\bf g}^2(\bfx)}{2m\rho(\bfx)} \Bigg ] =
{\cal F}_U \left[\rho\right] +{\cal F}_K \left[\rho,{\bf g}\right] 
\eea
In reaching the specific form of the so-called kinetic (momentum) part of the free energy functional, we use the isotropic form of the moment of inertia tensor $\rttensor{\kappa}_{ij}=\kappa_0\delta_{ij}$.
A microscopic calculation also reaches this form for the N particle system, starting with the Hamiltonian. \cite{langer-turski}

The form (\ref{tdgl-eq}) for the equation of motion also includes the so-called Smoluchowsky dynamics case involving a single collective variable, namely the density $(\rho({\bf x})$. In this case, the reversible part $V^a$ is identically zero, and we have only the dissipative part in the equation (see Eq. (\ref{DK-stat}) )for the dynamics. Finally, the free energy functional in this case is denoted as ${\cal F}_U[\rho]$ is identified with the Classical density functional theory expression for the free energy functional, with the corresponding chemical potential being obtained as 
$\mu({\bf x})={\delta{\cal F}_U}/{\delta\rho})$. The free energy ${\cal F}_U$ is obtained here as a functional of density only and involves a non-interacting (ideal gas) contribution ${\cal F}_{\rm id}$ and an interaction part ${\cal F}_{\rm in}$. 
\be \label{f-divide} {\cal F}_{U}={\cal F}_{\rm id}+{\cal F}_{\rm in}~~.\ee
The so-called direct correlation functions or thermodynamic structure functions, including angular correlations \cite{bagamu,lowen1}, are defined in terms of the functional derivatives of ${\cal F}_{\rm in}$.
\be \label{dir-cn} c^{(n)}({\bf x}_1,{\bf x}_2 ...)=\frac{{\delta^n} {\cal F}_{\rm in}}
{\delta\rho({\bf x}_1)\delta\rho({\bf x}_2).... )\delta\rho({\bf x}_n)}  ~~~.
\ee
For $n=2$, this is the usual Ornstein-Zernike \cite{ornstein} direct correlation function \cite{hansen,grey-gubbins}.

\section{DIscussion}

The rotational motion of the individual particles is included here in terms of the director ${\bf u}$ of constant length. The corresponding hydrodynamic description of the fluid with rotational degrees of freedom involves an extended set of continuum fields which includes, in addition to linear momentum density ${\bf g}(\bfx,t)$ the angular momentum density  $\bm{\ell}(\bfx,t)$.
The set of hydrodynamic fields $\{{\bf g}(\bfx,t),\bm{\ell}(\bfx,t)\}$ are averaged (over a non-equilibrium ensemble) of the corresponding microscopically defined densities $\{\hat{\bf g}(\bfx,t),\hat{\bm{\ell}}(\bfx,t)\}$. The so-called local equilibrium distribution is used as an approximation for calculating the averages over the non-equilibrium states. The local thermodynamic fields $\{\mu(\bfx,t),{\bf v}({\bf x},t),\bm{\omega}({\bf x},t)\}$  act as parameters defining the local equilibrium ensemble, and are not microscopically defined quantities. In strict  equilibrium these local thermodynamic fields are constants $\{\mu,{\bf v},\bm{\omega}\}$. Within this description, the equilibrium Gibbsian distribution may have a component $\exp\{-{\bf v}_0.{\bf P}-\bm{\omega}_0.{\bf L}\}$, for probability of states with centre of mass momentum ${\bf P}$ and angular momentum  about a lab fixed point ${\bf L}$. The thermodynamic parameters $\{{\bf v}_0,\bm{\omega}_0\}$ introduced through the Gibbsian distribution are, respectively, the linear speed of the centre of mass and the uniform angular speed of the whole system around the axis through the centre of mass. For ${\bf P}$ and ${\bf L}$ both zero, the $\bm{\omega}_0$ and ${\bf v}_0$ can also be taken to be zero. 

The stochastic PDEs for the time evolution of a set of continuum fields $\{\psi\}$ (say) are generally nonlinear and constitute the fluctuating nonlinear hydrodynamic (FNH) description \cite{spd-rmp}. In the present paper, we obtain these FNH equations, starting from the microscopic level description in terms of the equations of motion for the system's constituent particles. The equations of microscopic dynamics considered here are Langevin equations with frictional terms and noise. Hence, the particle dynamics is dissipative.
We show that the FNH equations obtained in V-B are the same as those obtained in Section V, using Poisson Brackets \cite{volvick} methods generally adopted for a Newtonian system. For the latter, the microscopic level dynamics is reversible and dissipation at the macroscopic equations are obtained from phenomenological considerations of positive entropy production in the system \cite{mpp}. The coarse-grained equations for FNH are very similar \cite{dm,munakata96,kawasaki-miyazima,dean,dawsan,velenich} for systems with very different microscopic dynamics. For microscopic dynamics being dissipative, expressed with stochastic equations, the dissipative terms in the FNH equations are linear in the fields. However, the momentum density equation's noise is multiplicative, as seen from the Eqns. (\ref{den-fnl})-(\ref{ell-fnl}) obtained in the present work. On the other hand, the microscopic dynamics are reversible for the Newtonian system, and the noise in the FNH equations is simple. Here, the dissipative term in the momentum equations has a $1/\rho$ nonlinearity \cite{dm-2009}. 

The FNH equations also involve the Free energy functional ${\cal F}[\psi]$ which represents the corresponding equilibrium state of the system since the stationary solution of the corresponding  Fokker-Planck equation ( for the probability ${\cal P}[\psi]$ of a state) is obtained as ${\cal P[\psi]}{\sim}\exp[-{\cal F(\psi)}]$. In the present work, we have shown how the two ${\cal F}$'s corresponding to the respective FNH descriptions with the sets $\{\rho,{\bf g},\bm{\ell}\}$ and $\{\rho\}$ only are related. This result is given in Eqn. (\ref{two-reln}).
For the even simpler cases of the two descriptions in terms of $\{\rho,{\bf g}\}$ and $\{\rho\}$, the corresponding relation is (\ref{one-rein}).
This relation (\ref{two-reln}) between the free energy functionals ${\cal F}[\rho,{\bf g},\bm{\ell}]$ and ${\cal F}_U[\rho]$ corresponding to the respective sets of fields 
$\{\rho,{\bf g},\bm{\ell}\}$ and $\{\rho\}$ is not reached by simply integrating out the extra fields from the Gibbsian distribution function corresponding to the extended set. Thus if we consider the equilibrium distribution $e^{-{\cal F}(\rho,{\bf g},\bm{\ell}]}$ and integrate out the fields $\{{\bf g},\bm{\ell}\}$ by doing the corresponding Gaussian (functional) integrals we do not reach the result  $\exp[-{\cal F}_U(\rho)]$).
The same applies to the equilibrium distributions
$e^{-{\cal F}[\rho,{\bf g}]}$ and $e^{{\cal F}_U[\rho]}$. By doing the Gaussian (functional) integrals with respect to ${\bf g}$ in $e^{-{\cal F}[\rho,{\bf g}]}$ we do not obtain the free energy ${\cal F}_U[\rho]$ in the exponent for the reduced distribution \cite{spd-book}. The so-called Smoluchowsky description with $\{\rho\}$  is reached from the Fokker-Planck description involving $\{\rho,{\bf g},\bm{\ell}\}$ fields (or $\{\rho,{\bf g}\}$) after the momentum densities ${\bf g},\bm{\ell}$ (or ${\bf g}$) are eliminated with the adiabatic or over-damping approximation described in earlier sections of this paper.

\section*{Acknowledgements}

A.Y. acknowledges the Ministry of Education, Culture, Sports, Science andTechnology of Japan for financial supportwith MEXT KAKENHI 24H01466. S. P. D acknowledges for financial support from the Anusandhan National Research Foundation (ANRF) through the JC Bose Fellowship Grant (JBR/2022/000004), of the Science \& Engineering Research Board (SERB), Department of Science and Technology, Govt. of India.

\newpage
\appendix{
    
    \section{Implications of multiplicative noise  in the director equation}
    \label{app1}
    
   To account for the orientational dynamics at the microscopic level, a director ${\bf u^\alpha}$ of fixed length is associated with the particle $\alpha$ ( where $\alpha=1, ...N$). The equation of motion for $u^\alpha_i(t)$ is obtained in Eqn.
  (\ref{e0-u}) as a stochastic PDE with multiplicative noise:
  \be
  \label{ae0-u} \gamma^{-1}_u \frac{d{\bf u}^\alpha}{dt}{=}
  \bm{\tau}^\alpha\times{\bf u}^\alpha +\bm{\zeta}^\alpha\times{\bf
      u}^\alpha\ee
  The corresponding stochastic process is obtained
  by integrating this equation over a time interval $\bar{t}=t$ to
  $\bar{t}=t+\Delta$. 
  \be
  \label{ae01} 
  \gamma^{-1}_u \Big \{{\bf u}^\alpha(t{+}\Delta){-}{\bf u}^\alpha(t)\Big \} {=}
  \int_t^{t+\Delta}d\bar{t}\{\bm{\tau}^\alpha(\bar{t}){\times}\bm{u}^\alpha(\bar{t})\}+
  \int_t^{t+\Delta}d\bar{t}\{\bm{\zeta}^\alpha(\bar{t})\times{\bf
      u}^\alpha(\bar{t})\}. \ee
  The stochastic part of the time evolution is denoted by the integral of the stochastic term  $\eps_{ijk}\zeta^\alpha_j{u}_k^\alpha$ over this time interval of
  length $\Delta{t}$. However, for multiplicative noise, this
  stochastic integral involves the dynamic variable
  ${u}_k^\alpha(t)$ itself. Depending on, if ${u}_k^\alpha(t)$ is
  evaluated at the beginning of the interval $\bar{t}=t$ or at the
  middle of the interval {\em, i.e.}, at $\bar{t}=t+\Delta/2$ we
  have what are termed as respectively the I\^{t}o and Stratonovich interpretations. In the Stratonovich interpretation, the stochastic part of  Eqn. (\ref{ae01}) is  calculated as follows :
  \bea \label{st-e1}&\ &\eps_{ijk}
  {u}_k^\alpha(t{+}\frac{\Delta}{2})
  \int_t^{t{+}\Delta}d\bar{t}\zeta^\alpha_j (\bar{t})= \eps_{ijk}\left [ {u}_k^\alpha(t){+}\int_{t}^{t{+}\frac{\Delta}{2}} d{t^\prime}\dot{u}_k^{\alpha}({t^\prime}) \right ] \int_t^{t{+}\Delta}d\bar{t}\zeta^\alpha_j (\bar{t}) \nonumber \\
  &=&\eps_{ijk}\left [{u}_k^\alpha(t) \int_t^{t{+}\Delta}d\bar{t}\zeta^\alpha_j (\bar{t})+
  \gamma_u{\eps_{klm}}
  \int_t^{t{+}\Delta}d\bar{t}\int_{t}^{t+\frac{\Delta}{2}}d{t^\prime}
  \zeta^\alpha_j(\bar{t})\zeta^\alpha_l (t^\prime)u^\alpha_m (t^\prime)\right ]~~. 
  \eea
  In writing the second equality, we replace the
  $\dot{u}^\alpha_k(t^\prime)$ by its stochastic part, which contributes at linear order and ignores the regular part in the higher order. 
  The average contribution of the second term is on the right-hand side
  of Eqn. (\ref{st-e1}) is of the same order as the first term. Averaging over the noise, the second term is on the right-hand side of Eqn. (\ref{st-e1}) obtain,
  \bea \label{st-e2}&&  
  \eps_{ijk}{u}_k^\alpha(t) \int_t^{t{+}\Delta}d\bar{t}\zeta^\alpha_j (\bar{t})+
  \gamma_u \eps_{ijk}\eps_{klm}
  \left \{  \int_t^{t{+}\Delta}d\bar{t}
  \int_{t}^{t+\frac{\Delta}{2}}d{t^\prime}
  \{2T\gamma_u^{-1}\}\delta_{jl}\delta(\bar{t}-t^\prime) \right \} u^\alpha_m(t^\prime)
  \nonumber \\
  &=&   \eps_{ijk}{u}_k^\alpha(t) \int_t^{t{+}\Delta}d\bar{t}\zeta^\alpha_j (\bar{t})
  -\eps_{ijk}\eps_{mjk}  \int_t^{t{+}\frac{\Delta}{2}}d\bar{t}(2T)u^\alpha_m(\bar{t})  \nonumber \\
  &=& \int_t^{t{+}\Delta}d\bar{t} {\Big [ \bm{\zeta}^\alpha(\bar{t})\times{\bf u}^\alpha(t){-} {2T}{\bf u}^\alpha(\bar{t}) \Big ]}_i{+}O(\Delta^2)~, \eea
  where we have used the property of Levi-Civita symbols that
  $\eps_{ijk}\eps_{kjm}=-\eps_{ijk}\eps_{mjk}=-2\delta_{im}$ in three dimensions.
  With this, the difference equation (\ref{st-e1})  for ${\bf u}^\alpha$ takes two different forms respectively for the Stratonovich and I\^{t}o interpretations of the multiplicative noise in the corresponding Langevin equation (\ref{e0-u})). For the I\"{t}o interpretation, we have the difference equation as:
  \be
  \label{ae-ito} 
  \gamma^{-1}_u \Big \{{\bf u}^\alpha(t{+}\Delta){-}{\bf u}^\alpha(t)\Big \}{=}
  \int_t^{t+\Delta}d\bar{t} \{\bm{\tau}^\alpha(\bar{t}){\times}\bm{u}^\alpha(\bar{t})\} +\Delta^{\frac{1}{2}}\bar{\bm{\zeta}}^\alpha(t)\times{\bf
      u}^\alpha(t)~.   \ee
  where we have defined the noise $\bar{\bm{\zeta}}^\alpha(t)$ as
  \be
  \label{av-noise}
  \bar{\bm{\zeta}}^\alpha(t){=}\Delta^{-\frac{1}{2}}
  \int_t^{t+\Delta}d\bar{t}\bm{\zeta}^\alpha(\bar{t})~~.
  \ee
  On the other hand, for the Stratonovich interpretation we obtain for the difference integral the result:
  \be
  \label{ae-stat} 
  \gamma^{-1}_u \Big \{{\bf u}^\alpha(t{+}\Delta){-}{\bf u}^\alpha(t)\Big \}{=}
  \int_t^{t+\Delta}d\bar{t} \left \{ \bm{\tau}^\alpha(\bar{t}){\times}\bm{u}^\alpha(\bar{t}){-}2T{\bf u}^\alpha(\bar{t}) 
  \right \}{+}\Delta^{\frac{1}{2}}\bar{\bm{\zeta}}^\alpha(t){\times}{\bf
      u}^\alpha(t)~.   \ee
  Using the result (\ref{tor-cor}) for the correlation of the noise $\bm{\zeta}^\alpha$ the correlation of the averaged noise $\bar{\bm{\zeta}}^\alpha$ is obtained as
  \be \label{av-tor-cor}
  <\bar{\bm{\zeta}}_i^\alpha(t)\bar{\zeta}_j^\nu(t')>=
  {2\kappa_0{T}}{\gamma_u}^{-1}
  \delta_{\alpha\nu}\delta_{ij}\delta_{t,t^\prime}~~. \ee
  Therefore, with the presence of multiplicative noise in the equation of motion for the director ${\bf u}$, which is introduced to account for the orientational motions in the microscopic level description of the dynamics of the particles, there, the time evolution is different, and this gets reflected in the coarse-grained equations of hydrodynamics describing the macroscopic level description.

\section{The Poisson Brackets}

The collective densities of number, linear momentum and angular momentum are respectively defined as follows:
\label{app2}
\bea \label{def1-rho} \hat{\rho}({\bf x},t)&=&
\sum_{\alpha=1}^N \delta({\bf r}-{\bf r}^\alpha(t)) 
\delta({\bf u}-{\bf u}^\alpha(t)) {\equiv} \sum_{\alpha=1}^N
\delta({\bf x}-{\bf x}^\alpha(t)) ~~, \\
\label{def1-g} \hat{\bf g}({\bf x},t) &=& \sum_{\alpha=1}^N
{\bf p}^\alpha(t) \delta({\bf x}-{\bf x}^\alpha(t))~, \\
\label{def1-l} \hat{\bm{\ell}}({\bf x},t)   &=& \sum_{\alpha} 
\bm{\ell}^\alpha(t) \delta({\bf x}-{\bf x}^\alpha(t)). \eea
The set of collective variables $\{\hat{\rho},\hat{\bf g},\hat{\bm{\ell}}\}$ defined above are microscopic quantities dependent on the phase space variables
$\{{\bf x}^\alpha,{\bf g}^\alpha,\bm{\ell}^\alpha\}$ for $\alpha=1,...N$.
The Poisson Brackets between the collective variables are obtained following the basic definition of Poisason Brackets given in Eqs. (\ref{pblu})-(\ref{pbrp}). We list below the results:
\bea
\label{rho-rho} 
\hat{Q}_{\rho\rho} (\bfx,\bfx^\prime){\equiv}\{\hat{\rho}(\bfx),\hat{\rho}(\bfx^\prime)\}&=& 0\\
\label{rho-g}
\hat{Q}_{\rho{g_i}} (\bfx,\bfx^\prime){\equiv}
\{\hat{\rho}(\bfx),\hat{g}_i(\bfx^\prime)\}&=& -\nabla^i_{\bf r}[\delta({\bf x}-{\bf x}^\prime)\hat{\rho}(\bfx)]\\
\label{rho-ell}
\hat{Q}_{\rho{\ell_i}}(\bfx,\bfx^\prime){\equiv}
\{\hat{\rho}(\bfx),\hat{\ell}_i(\bfx^\prime)\}&=&-\widetilde{\nabla}^i_{\bf u}[\delta({\bf x}-{\bf x}^\prime)\hat{\rho}(\bfx)] \\
\label{g-rho}
\hat{Q}_{{g_i}\rho} (\bfx,\bfx^\prime){\equiv}
\{\hat{g}_i(\bfx),\hat{\rho}(\bfx^\prime)\}&=& \nabla^i_{{\bf r}^\prime}[\delta({\bf x}-{\bf x}^\prime)\hat{\rho}(\bfx)] \\
\label{g-g}
\hat{Q}_{g_i{g_j}} (\bfx,\bfx^\prime){\equiv}
\{\hat{g}_i(\bfx),\hat{g}_j(\bfx^\prime)\}&=&   -\nabla^j_{\bf r}[\delta({\bf x}-{\bf x}^\prime)\hat{g}_i(\bfx)]{+}\nabla^i_{{\bf r}^\prime}[\delta({\bf x}-{\bf x}^\prime)\hat{g}_j(\bfx^\prime)]\\
\label{g-ell}
\hat{Q}_{g_i{\ell_j}} (\bfx,\bfx^\prime){\equiv}\{\hat{g}_i(\bfx),\hat{\ell}_j(\bfx^\prime)\}&=& -\widetilde{\nabla}^j_{\bf u}[\delta({\bf x}-{\bf x}^\prime)\hat{g}_i(\bfx)]
{+}{\nabla}^i_{{\bf r}^\prime}[\delta({\bf x}-{\bf x}^\prime)\hat{\ell}_j(\bfx^\prime)] \\
\label{ell-rho}
\hat{Q}_{{\ell_i}\rho} (\bfx,\bfx^\prime){\equiv}
\{\hat{\ell}_i(\bfx),\hat{\rho}(\bfx^\prime)\}&=& \widetilde{\nabla}^i_{{\bf u}^\prime}[\delta({\bf x}-{\bf x}^\prime)\hat{\rho}(\bfx^\prime)] \\
\label{ell-g}
\hat{Q}_{\ell_i{g_j}} (\bfx,\bfx^\prime){\equiv}\{\hat{\ell}_i(\bfx),\hat{g}_j(\bfx^\prime)\}&=& \widetilde{\nabla}^i_{\bf u^\prime}[\delta({\bf x}-{\bf x}^\prime)\hat{g}_j(\bfx^\prime)]
{-}{\nabla}^j _{\bf r}[\delta({\bf x}-{\bf x}^\prime)\hat{\ell}_i(\bfx)] \\
\label{ell-ell}
\hat{Q}_{\ell_i{\ell_j}}(\bfx,\bfx^\prime){\equiv}
\{\hat{\ell}_i(\bfx),\hat{\ell}_j(\bfx^\prime)\}&=& -\widetilde{\nabla}^j_{\bf u}[\delta({\bf x}-{\bf x}^\prime)\hat{\ell}_i(\bfx){+}\widetilde{\nabla}^i_{{\bf u}^\prime}[\delta({\bf x}-{\bf x}^\prime)\hat{\ell}_j(\bfx^\prime)]
\eea
For the coarse-grained variables or the continuum fields, the so called Poisson brackets $\{Q_{ab}\}$, for $\psi_a,\psi_b\in$$\{{\rho},{\bf g},\bm{\ell}\}$ are obtained from the corresponding microscopic quantity  $\hat{Q}_{ab}{\equiv}\{\hat{\psi}_a({\bf x}),\hat{\psi}_b({\bf x}^\prime)\}$ by replacing the hatted quantities in the expressions on the right hand sides of (\ref{rho-rho})-(\ref{ell-ell}),with their respective un-hatted counterparts. Using these results in expressions (\ref{v_rho})-(\ref{v_ell}) we obtain the  respective currents $V_\rho$, $V_i^g$, and $V_i^\ell$ in the FNH equations for the fields $\{{\rho},{\bf g},\bm{\ell}\}$.

The current term $V_\rho$ in the equation for mass density $\rho$ is obtained as,
\bea
\label{vrho}
V_\rho&\equiv&  \sum_{\phi=\rho,{\bf g},\bm{\ell}}
{\int d{\bfx}^\prime}Q_{\rho\phi}(\bfx,\bfx^\prime)
\frac{\delta{\cal F[\rho,{\bf g},\bm{\ell}]}}{\delta{\phi}(\bfx^\prime)} \nonumber \\
&=& - \int d\bfx^\prime 
Q_{\rho{g_i}} (\bfx,\bfx^\prime)\frac{\delta{\cal F[\rho,{\bf g},\bm{\ell}]}}{\delta{g_i}(\bfx^\prime)}
- \int d\bfx^\prime
Q_{\rho{\ell_i}} (\bfx,\bfx^\prime)\frac{\delta{\cal F[\rho,{\bf g},\bm{\ell}]}}{\delta{\ell_i}(\bfx^\prime)} \nonumber \\ 
&=& -\frac{1}{m}\nabla_{\bf r}^i{g_i}
-\frac{1}{\kappa_0}\widetilde{\nabla}_{\bf u}^i {\ell_i} ~~. \eea
Next,  the current term $V_g^i$ in the equation for momentum density $g_i$ we obtain,
\bea
\label{vmom}
V^i_{g}&\equiv&\sum_{\phi=\rho,{\bf g},\bm{\ell}}
{\int d{\bfx}^\prime}Q_{{g_i}\phi}(\bfx,\bfx^\prime)
\frac{\delta{\cal F[\rho,{\bf g},\bm{\ell}]}}{\delta{\phi}(\bfx^\prime,t)} \nonumber \\
&=& -\rho(\bfx,t)\nabla_{\bf r}\frac{\delta{\cal F}}{\delta{\rho}(\bfx,t)}-
\nabla_{\bf r}^j\left\{ g_i(\bfx,t)\frac{\delta{\cal F}} {\delta{g_j}(\bfx,t)} \right\}   
+ g_j(\bfx,t)\nabla_{\bf r}^i\left\{\frac{\delta{\cal F}} {\delta{g_j}(\bfx,t)} \right\}\nonumber \\
&-&   \nabla_{\bf r}^j\left\{ g_i(\bfx,t)
\frac{\delta{\cal F}} {\delta{\ell_j}(\bfx,t)} \right\} 
+{\ell_j}(\bfx,t) {\nabla}_{\bf r}^i\left \{
\frac{\delta{\cal F}}{\delta{\ell}_j(\bfx,t)} \right \}
\nonumber \\  
&=& -\rho\nabla_{\bf r}\frac{\delta{\cal F}_U}{\delta{\rho}}{-}
\nabla_{\bf r}^j\left [ \frac{g_i{g_j}}{\rho} \right ]{-}\nabla_{\bf r}^j\left [
\frac{g_i{\ell_j}}{\rho} \right ]+{\cal I}^{(g)}_i ~~. 	  
\eea
where the integral ${\cal I}^{(g)}_i$ is obtained as 
\bea
{\cal I}_i^{(g)}&=&-\rho(\bfx)} \widetilde{\nabla}_{\bf u}^i
\frac{\delta{\cal F}_K^{(g)}}{\delta{\rho(\bfx )}}{+}\int{d}{\bfx}^\prime
\widetilde{\nabla}_{{\bf u}^\prime}^i[\delta({\bf x}-{\bfx}^\prime){g}_j(\bfx^\prime)]\frac{\delta{\cal F}}{\delta{g_j(\bfx^\prime)}}
{\nonumber \\
&=&{-}\left (\frac{g^2}{\rho^2}\widetilde{\nabla}_{\bf u}^i\rho{-}
\frac{g_j}{\rho}\widetilde{\nabla}_{\bf u}^i{g_j}\right ){-}
\widetilde{\nabla}_{\bf u}^i\left ({\frac{g_ig_j}{\rho}}\right )
{+}\frac{g_j}{\rho}\widetilde{\nabla}_{\bf u}^i{g_j} = 0
\eea
Finnally, the current term in the equation for angular momentum density $\ell_i$ we obtain,
\bea
\label{vamom}
V^i_{\ell}&\equiv&\sum_{\phi=\rho,{\bf g},\bm{\ell}}
{\int d{\bfx}^\prime}Q_{{\ell_i}\phi}(\bfx,\bfx^\prime)
\frac{\delta{\cal F[\rho,{\bf g},\bm{\ell}]}}{\delta{\phi}(\bfx^\prime)} \nonumber \\
&=&\Big \{{-}{\rho(\bfx)} \widetilde{\nabla}_{\bf u}^i
\frac{\delta{\cal F}}{\delta{\rho(\bfx)}}{-}\nabla_{\bf r}^j\frac{\delta{\cal F}}{\delta{g_j(\bfx)}}	
{\ell_i(\bfx)}{+}\int{d}{\bfx}^\prime
\widetilde{\nabla}_{{\bf u}^\prime}^i[\delta({\bf x}-{\bfx}^\prime)\hat{g}_j(\bfx^\prime)]\frac{\delta{\cal F}}{\delta{g_j(\bfx^\prime)}} \nonumber \\
&-&\widetilde{\nabla}_{\bf u}^j \Big \{ \frac{\delta{\cal F}}{\delta{\ell}_j(\bfx)}{\ell}_i(\bfx)
\Big \}{+}\int{d}{\bfx}^\prime
\widetilde{\nabla}^i_{{\bf u}^\prime}[\delta({\bf x}-{\bf x}^\prime){\ell}_j(\bfx^\prime)]
\frac{\delta{\cal F}}{\delta{\ell}_j(\bfx^\prime)}
\nonumber \\
&=& 
-{\rho}\widetilde{\nabla}_{\bf u}^i
\frac{\delta{\cal F}_U}{\delta{\rho}}-\nabla_{\bf r}^j\Big [ 
\frac{{\ell_i}{g_j}}{\rho} \Big ]
-\widetilde{\nabla}_{\bf u}^j \Big [ \frac{{\ell_i}{\ell_j}}{\rho} \Big ]
+{\cal I}^{(\ell)}_i ~~.
\eea
where the integral ${\cal I}^{(\ell)}_i$ is obtained as 
\bea
{\cal I}_l^{(\ell)}&=&{-}{\rho(\bfx)} \widetilde{\nabla}_{\bf u}^i
\frac{\delta{\cal F}_K^{(\ell)}}{\delta{\rho(\bfx)}}{+}
\int{d}{\bfx}^\prime
\widetilde{\nabla}_{{\bf u}^\prime}^i[\delta({\bf x}-{\bfx}^\prime){\ell}_j(\bfx^\prime)]
\frac{\delta{\cal F}}{\delta{\ell_j(\bfx^\prime)}}\nonumber \\
&=&{-}\left (\frac{\ell^2}{\rho^2}\widetilde{\nabla}_{\bf u}^i\rho{-}
\frac{\ell_j}{\rho}\widetilde{\nabla}_{\bf u}^i{\ell_j}\right ){-}
\widetilde{\nabla}_{\bf u}^i\left ({\frac{\ell_i{\ell_j}}{\rho}}\right )
{+}\frac{\ell_j}{\rho}\widetilde{\nabla}_{\bf u}^i{\ell_j}=0
\eea
Using the results (\ref{vrho}), (\ref{vmom}), and (\ref{vamom}) we obtain the corresponding expressions (\ref{v_rho}), (\ref{v_g}), and (\ref{v_ell}) for the currents appearing in the FNH equations.

\end{document}